\documentclass[preprint2,longabstract]{aastex}

\newcommand{\eg}{{\frenchspacing\it e.g.}}

\newcommand{\etal}{{\frenchspacing\it et al. }}
\newcommand{\kms}{ km s$^{-1}$ }
\newcommand{\lsim}{\hbox{ \rlap{\raise 0.425ex\hbox{$<$}}\lower
0.65ex\hbox{$\sim$} }}
\newcommand{\gsim}{\hbox{ \rlap{\raise 0.425ex\hbox{$>$}}\lower
0.65ex\hbox{$\sim$} }}

\shorttitle{X-ray Galaxy Clusters in NoSOCS}
\shortauthors{Lopes et al.}

\begin{document}

\title{X-ray Galaxy Clusters in NoSOCS: Substructure and the Correlation of Optical and X-ray Properties}

\author{P.A.A. Lopes\altaffilmark{1}, R.R. de Carvalho, H.V. Capelato}

\affil{Instituto Nacional de Pesquisas Espaciais - Divis\~ao de 
Astrof{\' \i}sica, Avenida dos Astronautas, 1758, S\~ao Jos\'e dos Campos, SP 12227-010, Brasil}

\author{R.R. Gal}

\affil{UC Davis, Dept. of Physics, One Shields Ave, Davis, CA 95616}

\author{S.G. Djorgovski}

\affil{Palomar Observatory, Caltech, MC 105-24, Pasadena, CA 91125}

\author{R.J. Brunner}

\affil{University of Illinois, Dept. of Astronomy, 1002 W. Green St., Urbana, IL 61801}

\author{S.C. Odewahn}

\affil{Hobby-Eberly Telescope, HC75 - Box 1337-10, Ft. Davis, TX 79734-5015}

\author{and}

\author{A.A. Mahabal}

\affil{Palomar Observatory, Caltech, MC 105-24, Pasadena, CA 91125}

\altaffiltext{1}{Email: paal@das.inpe.br}

\begin{abstract}
We present a comparison of optical and X-ray properties of galaxy
clusters in the northern sky, using literature data from BAX and
optically selected clusters in DPOSS. We determine the recovery rate
of X-ray detected clusters in the optical as a function of richness,
redshift and X-ray luminosity, showing that the missed clusters are
typically low contrast systems when observed optically (either poor or
at high redshifts). We employ four different statistical tests to test
for the presence of substructure using optical two-dimensional
data. We find that approximately $35 \%$ of the clusters show strong
signs of substructure in the optical. However, the results are
test-dependent, with variations also due to the magnitude range and
radius utilized. We have also performed a comparison of X-ray
luminosity and temperature with optical galaxy counts (richness). We
find that the slope and scatter of the relations between richness and
the X-ray properties are heavily dependent on the density contrast of
the clusters. The selection of substructure-free systems does not
improve the correlation between X-ray luminosity and richness, but
this comparison also shows much larger scatter than one obtained using
the X-ray temperature.  In the latter case, the sample is
significantly reduced because temperature measurements are available
only for the most massive (and thus high contrast) systems. However,
the comparison between temperature and richness is very sensitive to
the exclusion of clusters showing signs of substructure. The
correlation of X-ray luminosity and richness is based on the largest
sample to date ($\sim$ 750 clusters), while tests involving
temperature use a similar number of objects as previous
works ($\lsim$100). The results presented here are in good agreement
with existing literature.
\end{abstract}

\keywords{galaxies: clusters: general -- X-rays: galaxies -- catalogs}

\section{Introduction}

In most scenarios of structure formation, galaxy clusters are the
largest and latest objects to form under the influence of their own
gravity. Reliable measurements of cluster masses allow us to study the
evolution of the large scale structure through the cluster mass
function and its time evolution.  Several studies have examined the
evolution experienced by clusters, using the results to provide
estimates of important cosmological parameters, such as $\Omega_m$ and
$\sigma_8$ \citep{eke98, mat98, kit97, bah97, bah03, car97, don98,
rei99, bla98, pos02}.  The main difficulty faced by these studies lies
in determining the cluster masses.  A variety of techniques can be used
for this measurement, including dynamical methods (velocity
dispersions), measuring the temperature of the intra-cluster gas,
observations of the Sunyaev$-$Zeldovich effect, and weak gravitational
lensing. All of these methods are observationally
expensive, which render large scale structure studies based on extensive
samples impractical.

An alternative method to obtain mass estimates for a large sample of
clusters is to connect the cluster mass to an easily observable
parameter, such as X-ray luminosity (L$_X$) or the number of
constituent galaxies (richness). The tightest mass-observable
relations involve X-ray temperatures (T$_X$) and spectroscopically
measured velocity dispersions ($\sigma_v$).  Unfortunately, large
samples of clusters with either or both of these parameters measured
do not exist currently. The X-ray luminosity (L$_X$) and optical
richness (or luminosity, L$_{opt}$) also correlate well with mass,
although with larger scatter than relations with T$_X$ and
$\sigma_v$. The precise measurement and calibration of these relations
are essential for reliable mass function estimates. For instance,
underestimation of the scatter in the mass$-$observable relation can
result in overestimation of $\sigma_8$ \citep{voi05}. Furthermore, the
correlation of mass to properties determined at different wavelengths
can lead to contradictory results, which may be associated with the
selection function of the cluster catalogs constructed in different
regimes, or with physical processes and evolution within the cluster
populations.

Thus, understanding the systematics present in optical and X-ray
surveys, and the way they complement each other, are necessary
precursors to studies of cluster evolution and are even more important
for cosmological tests that require statistically robust cluster
samples. Some of these issues have been addressed in recent
literature, either by comparing X-ray and optical cluster catalogs or
conducting joint X-ray/optical surveys of galaxy clusters
\citep{don01, don02, gil01, gil04, yee03, bas04, pop04, pop05, smi05}. A very
important by-product of such comparisons is the selection and
follow-up study of unusual clusters, such as those that are optically rich but X-ray underluminous \citep{gil01, gil04,lub04}.

As mentioned above, an important issue concerning these scaling
relations is the precision to which they can be established and which
observational biases contribute to increasing their scatter. Information
on the distribution of galaxies, hot gas and dark matter within
clusters can play an important role in understanding this
scatter. The diversity in the dynamical states of galaxy clusters may
be associated with the uncertainties in the determination of the
scaling relations. The presence of substructure is a clear sign of
incomplete relaxation in a cluster, and can be estimated from the X-ray
emission from the intra-cluster medium as well as the distribution (one, two
or three dimensional) of cluster galaxies \citep{mat99, kol01, smi05}.
The dynamical state of galaxy clusters is also strongly related to the
underlying cosmology, so that the evolution of cluster substructure
with look-back time is a powerful tool for constraining
cosmological parameters \citep{moh95}.

A few recent works have investigated the possibility that cluster
substructure adds scatter to the observed scaling relations, but the
results have been ambiguous. While \citet{smi05} find that the
observational scaling relations of clusters behave differently for
relaxed and unrelaxed clusters, \citet{oh06} find the opposite (based
both on simulations and observational data). Some of these differences
may be due to the relatively small sample size or to the inconsistent
techniques for measuring substructure.

The main goal of this paper is the establishment of scaling relations
for galaxy clusters through the comparison of optical and X-ray
properties (N$_{gals}$, L$_X$, T$_X$). We also want to examine which
factors contribute to the observed scatter in these relations. We
first investigate the recovery rate of X-ray emitting galaxy clusters
in an optical survey, searching specifically for X-ray detected
clusters which could be missed in the optical. We also conduct a
substructure study based solely on the photometric data. Finally, we
compare the optical and X-ray properties of galaxy clusters,
investigating the dependence of the results with the contrast cut of
the sample used, the centroid adopted for the richness measures and
the amount of substructure. We limit this study to photometric and
X-ray parameters and investigate the systematics in these
comparisons, postponing to a future work the extension of the cluster
scaling relations for other parameters, such as $R_{200}$, $\sigma_v$
and especially mass.

We use a list of X-ray observed galaxy clusters from BAX
(\url{http://bax.ast.obs-mip.fr/}) as our basis. The catalog initially
comprises 914 galaxy clusters in the northern hemisphere, spanning the
redshift range $0.05 \le z \le 0.40$. We trim this list to the same
area covered by the Northern Sky Optical Cluster Survey (NoSOCS) in
order to compare the two catalogs. The final X-ray list contains 792
clusters, of which 638 ($81\%$) are detected in the optical.

The remainder of this paper is organized as follows. In $\S$2 we
describe the X-ray data, while in $\S$3 we provide details about the
NoSOCS catalog.  The BAX and NoSOCS comparison is discussed in $\S$4,
while the substructure analysis is presented in $\S$5. In $\S$6 we
investigate the correlation between optical and X-ray properties,
namely N$_{gals}$, L$_X$ and T$_X$. The X-ray temperature-luminosity
relation is also determined. In particular we investigate the impact
of substructure and the cluster contrast on the scatter of these
relations. We summarize our results in $\S$7. We use H$_0 = 100$ $h$
\kms Mpc$^{-1}$, except where the $h-$dependence is explicitly given,
and $q_0 = 0.5$ ($h$ is fixed to 0.5 only for converting magnitudes,
but not for computing angular sizes). In order to be in agreement with
BAX (see \url{http://bax.ast.obs-mip.fr/html/help/bax\_help.html} for
notes on the BAX data) we decided not to adopt the currently favored
cosmology with $h = 0.70$, $\Omega_m= 0.30$ and $\Omega_{\Lambda} =
0.70$.  This choice has no significant effect on the results for the
scaling relations (namely slope, intercept and scatter). The only
parameters that might be affected by changing cosmology are those
associated with evolution, which is not expected to be significant in
the current work due to the small redshift range probed.

\section{X-ray Data}

We search for optical counterparts of X-ray observed clusters selected
from BAX, {\it Base de Donn\'ees Amas de Galaxies X}
(\url{http://bax.ast.obs-mip.fr/}). BAX is an on line research
database containing information on all galaxy clusters with X-ray
observations to date.  BAX contains all galaxy clusters confirmed as
X-ray sources, but is not restricted to objects first detected in
X-rays, and includes optically-selected clusters (such as Abell
clusters) with X-ray follow-up.  The database allows us to retrieve
each cluster's coordinates and redshift; various X-ray observational
measurements, namely flux (F$_{X}$), luminosity (L$_{X}$) and
temperature (T$_{X}$); and a set of bibliographical references. The
cluster coordinates and redshift are automatically generated from the
NASA/IPAC Extragalactic Database (NED\footnote{\url{http://nedwww.ipac.caltech.edu/}}). 
F$_{X}$ and L$_{X}$ are in the
ROSAT [0.1-2.4 keV] band. F$_{X}$ and L$_{X}$ units are $10^{-12}$
$ergs$ $s^{-1}$ $cm^{-1}$ and $10^{44}$ $ergs$ $s^{-1}$,
respectively. Only BAX {\it canonical} values of F$_{X}$, L$_{X}$ and
T$_{X}$ are retrieved. These are values chosen among the most precise
and/or latest published measurements, meaning that if there are
multiple references for a given cluster, the most precise and/or
recent is kept. The BAX output is given for H$_0$ = 50 km s$^{-1}$
Mpc$^{-1}$ and ${\Omega_m} =$ 1.0. At the time of our query, BAX contained
information on 1851 groups and clusters of galaxies, with 1656
clusters having luminosity measurements and 463 clusters
with temperature measurements.

We performed a multi-criterion search with the following constraints:
$-5^\circ \le \delta \le 90^\circ$ and $0.05 \le z \le 0.40$. These
limits ensure overlap with our optical galaxy cluster catalog
(NoSOCS), which is based on DPOSS ($\delta > -2.5^\circ$ and $z \lsim
0.3$). The query was performed on 2005 April 12, resulting in a list of
914 galaxy clusters.  We then compared the X-ray centers of these
clusters to all the plate limits used for NoSOCS, keeping only those
that fall within these limits. We found 800 X-ray clusters within the
NoSOCS limits, from which we eliminated eight that overlapped with bad
areas from our survey (excised regions due to saturated objects such as bright
stars). Of these 792 remaining systems, 638 have an optical
counterpart in our catalog.  From these 638 common clusters, 620 have 
measured X-ray luminosities (with two others discarded due to problems measuring the optical richness), while 101 have  X-ray
temperatures determined. More details on the recovery rate of X-ray
galaxy clusters by DPOSS are provided in section $\S$4. The comparison of
optical and X-ray properties, shown in section $\S$6, is based on this
list of 618 common clusters with available L$_X$ and N$_{gals}$, or
the subset of 101 clusters for which temperatures are available.

\section{Optical Data}

The X-ray galaxy cluster catalog derived from BAX is compared to the
Northern Sky Optical Cluster Survey (NoSOCS, Gal et al. 2003, 2006;
papers II and III, respectively), which is a catalog of galaxy
clusters constructed from the digitized version of the Second Palomar
Observatory Sky Survey (POSS-II; DPOSS). NoSOCS is derived from
high-latitude fields $|b| > 30^\circ$, covering $\sim 11000$ deg$^2$
and containing $\sim 16000$ cluster candidates. Richness and redshift
estimates are provided for nearly all clusters; the richness measure
(denoted N$_{gals}$) is described in detail in \citet{gal03} and in
the next section. The median redshift of the sample is $z_{med} =$
0.16, with a median richness of N$_{gals,med} =$ 31.  From the
selection functions presented in papers II and III we expect to find
rich clusters (N$_{gals} \sim 100$) out to $z \sim 0.3$. This is the
largest galaxy cluster catalog available to the present date. The
NoSOCS catalog and other DPOSS products can be found at
\url{http://dposs.ncsa.uiuc.edu/}.

\section{Richness and the BAX -- NoSOCS Comparison}

Due to differences between the photometric redshift estimates from
NoSOCS and the spectroscopic redshifts provided by BAX, we adopt the
BAX redshift for all analysis involving clusters common to the two
datasets. We note that the scatter in the comparison of NoSOCS and BAX
redshifts is comparable to the redshift accuracy estimated by
\citet{gal03}.  We utilize the BAX redshifts measures as they are
generally more accurate (as they are typically spectroscopic), and the
richness measure and its accuracy are sensitive to the cluster
centroid, radius and redshift adopted. This choice is corroborated by
the fact that the comparisons of optical and X-ray properties show
reduced scatter when using the BAX redshifts (described in Sections 4.1
and 6).  The improved correlations are expected, since we now compute
optical richnesses with the same redshifts used for calculating the
X-ray properties (L$_X$ and T$_X$ taken directly from BAX). In $\S$4.1
we detail the richness estimator, investigating its dependence on the
cluster centroid, redshift and aperture used. Then, in $\S$4.2 we show
the results -- in terms of recovery rate -- of the comparison between
the BAX clusters and the NoSOCS. We begin by searching for possible optical
counterparts in NoSOCS for each of the 792 BAX clusters, using a
matching radius of 1.50 h$^{-1}$ Mpc and keeping only the closest
match to each BAX entry. The total number of recovered systems is 638.

\begin{figure*}[!ht]
\centering
\includegraphics[totalheight=5.0in, scale=0.6, trim= 0 1.0in 0 0]{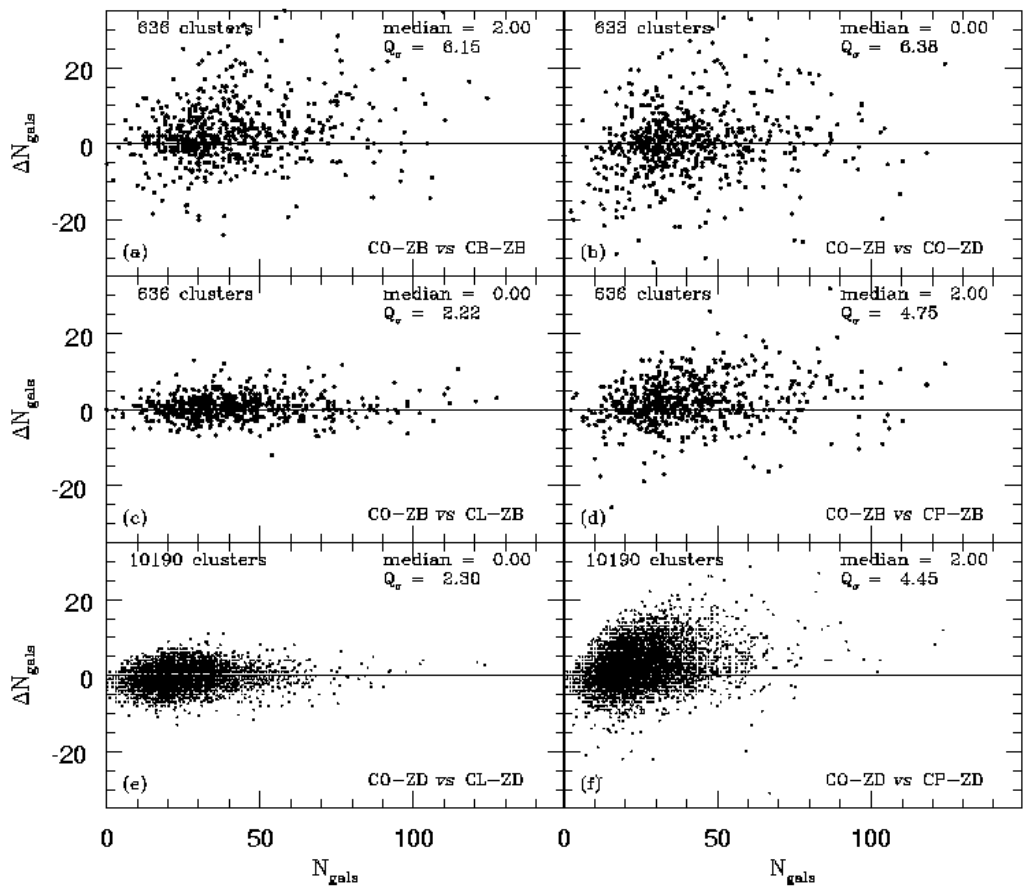}
\caption{Richness residual for different cluster centroids and adopted
redshifts. The labels in the lower right of each figure have the following 
meaning: ``C'' stands for coordinates and ``Z'' for redshifts; ``O'' is
the original cluster coordinate; ``B'' 
is the BAX centroid or redshift;
``L'' is  the luminosity weighted coordinate; ``P'' is the centroid given by the maximum density peak;
and ``D'' means the photometric redshift from DPOSS. Richness is compared 
for the following cases: ({\it a}) cluster original center {\it vs.} BAX centroid, 
both using the BAX redshift; ({\it b}) using the DPOSS
redshift {\it vs} the BAX redshift (both at the original coordinates); ({\it c}) original center {\it vs} the luminosity weighted 
coordinate (both using the BAX redshift); ({\it d}) original 
center {\it vs} the density peak centroid (both using the BAX redshift); ({\it e}) same as panel (c), but for the 10190
clusters used to estimate substructure; ({\it f}) same as panel (d), but for
all 10190 clusters in the NoSOCS sample. Panels (e) and (f) use the DPOSS photometric redshift by necessity.
\label{fig6_01}}
\end{figure*}

\subsection{Systematic Effects on Richness Estimation}

Measurement of the optical richness of a cluster 
depends on the cluster centroid, radius adopted, and the redshift. 
Our richness measure $N_{gals}$
is defined as the number of galaxies at 
$m^*_r - 1 \le m_r \le m^*_r + 2$ within a given aperture, where $m^*_r$ is the
characteristic apparent magnitude of the cluster luminosity function. For 
this work we assume a universal Schecter luminosity function, with 
parameters given by \citet{pao01}. The characteristic magnitude is 
$M^*_r = -22.17$, while $\alpha = -1.1$. As explained below, we 
find the optimal aperture to be 0.50 h$^{-1}$ Mpc. We have considered no 
evolution in the characteristic magnitude (m*). k$-$corrections are applied
to the magnitude of each galaxy, as explained in items (1) and (3) below.
These corrections are obtained trhough the convolution of spectral energy 
distributions from \citet{col80} with the DPOSS $r$ filter. The steps taken 
to estimate richness are:

\begin{enumerate}
\item We use the redshift of each cluster to convert the
characteristic magnitude to an apparent magnitude ($m^*_r$) and to
calculate the apparent radius (in seconds of arc) for a fixed aperture
of 0.50 h$^{-1}$ Mpc. If a cluster lies closer than 0.50 h$^{-1}$ Mpc
to a plate border we do not calculate its richness. The
redshift is also used to compute the k$-$correction values typical of
elliptical and late-type galaxies (Sbc) at the cluster redshift. These
are named ``ke'' and ``ks'', respectively. Because we only
want to count galaxies with $m^*_r - 1 \le m_r \le m^*_r + 2$, we
select all galaxies within 0.50 h$^{-1}$ Mpc of the cluster center at
$m^*_r - 1 + ks$ $ \le m_r \le$ $m^*_r + 2 + ke$.  The
k$-$corrections are applied to individual galaxies at a later stage,
so these limits guarantee that we select all galaxies that can fall within
$m^*_r - 1 \le m_r \le m^*_r + 2$. The number of galaxies selected in
the cluster region is N$_{clu}$. If the low ($m^*_r - 1 +
ks$) or high ($m^*_r + 2 + ke$) magnitude limits fall outside the survey
observational limits ($15.0 \le m_r \le 20.0$), we apply a correction to 
the richness measure (explained below).

\item We estimate the background contribution locally. We randomly
select twenty background boxes (1200$'' \times$ 1200$''$) within an
annulus ranging from 2.25 h$^{-1}$ Mpc to 2.25 h$^{-1}$ Mpc +
1.3$^{\circ}$ of the cluster center.  Galaxies are selected within the
same magnitude range as used for computing N$_{clu}$. We re-select boxes
that overlap with an excised area of the survey (due to bright stars,
for instance). The background counts of each box are scaled to the
same area as the cluster, and the median counts (from all 20 boxes) gives
the background estimate (N$_{bkg}$). We adopt the interquartile range
(IQR, which is the range between the first and third quartiles) as a
measure of the error in N$_{bkg}$, which we term
$Q_{\sigma_{bkg}}$ (for normally distributed data IQR = 1.35 $\times
\sigma$, where $\sigma$ is the standard deviation). The background
corrected cluster counts (N$_{clu}$ - N$_{bkg}$) is called N$_{corr}$.

\item Next, a bootstrap procedure is used to statistically apply
k$-$corrections to the galaxy populations in each cluster. In each of
100 iterations, we randomly select N$_{corr}$ galaxies from those
falling in the cluster region (N$_{clu}$). We then apply a k$-$correction to
the magnitude of each galaxy. An elliptical k$-$correction is applied to
80$ \%$ of the N$_{corr}$ galaxies, while an Sbc k$-$correction is
applied to the remaining 20$ \%$. Finally, we use these k$-$corrected
magnitudes to count the number of galaxies at $m^*_r - 1 \le m_r \le
m^*_r + 2$. The final richness estimate N$_{gals}$ is given by the median
counts from the 100 iterations. The richness error from the bootstrap
procedure alone is given by $Q_{\sigma_{boot}}$ = IQR. The total richness error
is the combination of this error and the background contribution, so
that $Q_{\sigma_{N_{gals}}}$ = $\sqrt{Q_{\sigma_{boot}}^2 +
Q_{\sigma_{bkg}}^2}$.

\item If the cluster is too nearby or too distant,
either the bright ($m^*_r - 1 + ks$) or faint ($m^*_r + 2 +
ke$) magnitude limit, respectively, will exceed one of the survey
limits ($15.0 \le m_r \le 20.0$). In this case we need to apply one of
the following correction factors to
the richness estimate:

$$\gamma_1 = {\int_{m_r^*-1}^{m_r^*+2} \Phi(m)dm \over
\int_{15}^{m_r^*+2} \Phi(m)dm} \eqno (1)$$
$$\gamma_2 = {\int_{m_r^*-1}^{m_r^*+2} \Phi(m)dm \over
\int_{m_r^*-1}^{20} \Phi(m)dm} \eqno (2)$$ We call $\gamma_1$ and
$\gamma_2$ the low and high magnitude limit correction factors. Whenever 
necessary, one of the above factors is multiplied by N$_{gals}$.
Because the optical data span five magnitudes it is
impossible to simultaneously exceed both the bright and faint end.
\end{enumerate}

An estimate of the cluster contrast is also obtained along with the
richness.  The contrast is defined as the ratio between the
number of galaxies in the cluster region having $m^*_r - 1 + ks$ $
\le m_r \le$ $m^*_r + 2 + ke$ (N$_{clu}$), and the background
error $Q_{\sigma_{bkg}}$ (estimated within the same magnitude range).

We utilize this richness measure to examine its dependence on the
cluster centroid, radius and redshift. In Figure 1 we show the
residual between different richness estimates obtained when adopting
different centroids and redshifts. In the lower right corner of each
panel, we encode the coordinates ``C'' and redshifts ``Z'' used as follows:
\begin{itemize}
\item CO: the original cluster position from NoSOCS, which is the location provided by SExtractor in an adaptively smoothed density map;
\item CL: a luminosity weighted coordinate, where each galaxy position
within the cluster aperture is weighted by the galaxy's luminosity and
a new center is computed;
\item CB: the cluster center cataloged in BAX;
\item CP: the coordinate of the maximum density peak;
\item ZD: the photometric redshift from the DPOSS data;
\item ZB: the redshift (usually but  
not always spectroscopic) cataloged in BAX.
\end{itemize}
In the two lower panels we show the results
obtained with all 10190 clusters used to estimate substructure
(section 5), while in the upper four panels the analysis is restricted
to the 636 clusters common to BAX and DPOSS.

\begin{figure}[!ht]
\includegraphics[totalheight=5.0in, scale=0.6, trim= 1.6in 0 0 0]{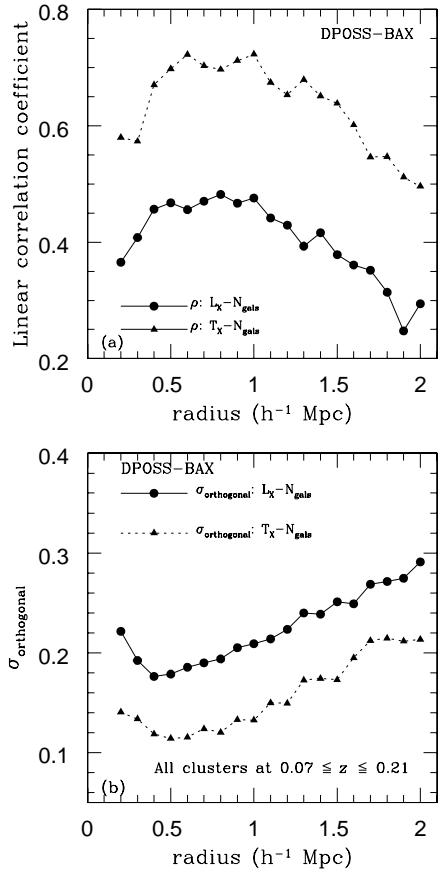}
\caption{{\it Top}: dependence of the linear correlation
coefficient ($\rho$) for the the L$_X-$N$_{gals}$ (solid line)
and T$_X-$N$_{gals}$ relations (dotted line) as a function of the radius;
{\it Bottom}: variation of the orthogonal scatter of the L$_X-$N$_{gals}$ (solid line)
and T$_X-$N$_{gals}$ relations (dotted line) as a function of the radius
used for calculating richness.
\label{fig3}}
\end{figure}

\begin{figure}[!hb]
\centering
\includegraphics[totalheight=2.7in, scale=0.6, trim= 0.6in 2in 0 0]{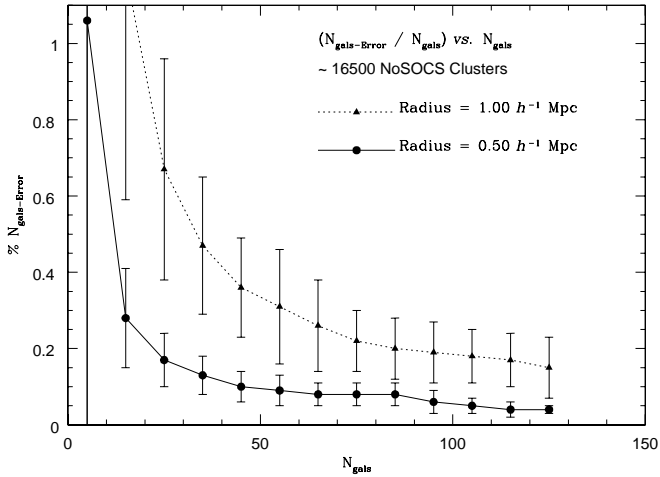}
\caption{Relative richness error as a function of richness for the full
NoSOCS sample ($\sim$ 16500 clusters). The solid line is the
variation of relative richness error using a 0.50 $h^{-1}$ Mpc
aperture for computing richness. The variation when adopting a radius
of 1.00 $h^{-1}$ Mpc is shown by the dotted line.  
\label{fig4}}
\end{figure}

\begin{deluxetable}{ccccc}
\tabletypesize{\footnotesize}
\tablecolumns{5}
\tablewidth{0pc}
\tablecaption{Scatter and number of objects in the optical {\it vs} 
X-ray relations for different richness estimates}
\tablehead{
\colhead{Set} & \colhead{L$_X-$N$_{gals}$} & \colhead{L$_X-$N$_{gals}$} &
\colhead{T$_X-$N$_{gals}$} & \colhead{T$_X-$N$_{gals}$} \\
\colhead{} & \colhead{All clusters} & \colhead{$\beta_{signif > 0.05}$} & 
\colhead{All clusters} & \colhead{$\beta_{signif > 0.05}$}
}
\startdata
CO-ZB & 0.172; 430 & 0.171; 271 & 0.112; 53 & 0.041; 23 \\
CB-ZB & 0.184; 418 & 0.174; 233 & 0.119; 53 & 0.061; 26 \\ 
CO-ZD & 0.179; 424 & 0.175; 266 & 0.111; 48 & 0.111; 29 \\ 
CL-ZB &  0.174; 432 & 0.174; 274 & 0.120; 53 & 0.062; 28 \\
CP-ZB &  0.174; 426 & 0.175; 260 & 0.120; 53 & 0.065; 26 \\
\enddata
\end{deluxetable}

\begin{figure}[!hb]
\centering
\includegraphics[totalheight=3.0in, scale=0.6, trim= 0.6in 0.7in 0 0]{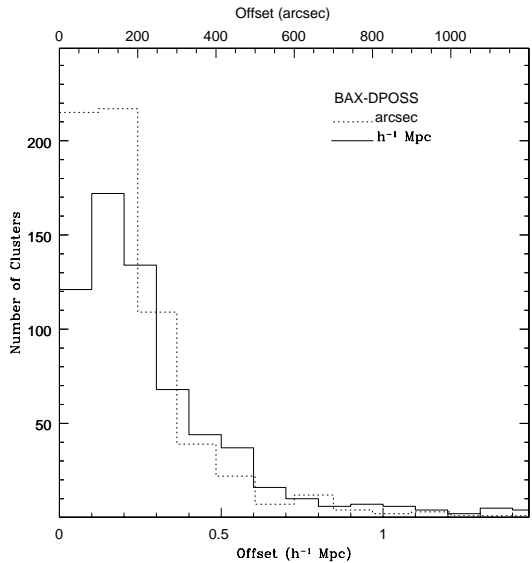}
\caption{Offset distribution between BAX and NoSOCS clusters in 
h$^{-1}$ Mpc (solid line) and arcseconds (dotted line).\label{fig1}}
\end{figure}

From inspection of this figure it is clear that the richnesses
estimated with the X-ray centroids are underestimated compared to
those obtained with the optical centers (panel a). This is likely due
to the optical center being a better choice than the X-ray center to
trace the distribution of galaxies within the cluster. If we restrict
the study to clusters with small optical--X-ray centroid offsets
($\lsim$ 0.50 h$^{-1}$ Mpc) the results are very similar, although a
small trend for higher richness values computed with the optical
centers is expected. When adopting the same center (optical) and using
different redshifts (from BAX and DPOSS) there is no obvious trend
(the median residual is zero), but the scatter is the largest of all
the comparisons (panel b). This implies that there is no systematic
difference between the photometric redshifts and those from BAX. The
scatter arises from a shift in sampling the luminosity function of
each cluster and a change in radius when estimating richness. Adopting
either the original or the luminosity weighted coordinates does not
appreciably affect the resultant richnesses (panels c and e), as the
two centers are typically similar. On the other hand, the richness
estimate obtained with the density peak center is underestimated in
comparison to the other optical centroids (panels d and f). This may
be due to the presence of substructure (or projection effects), as a
cluster showing more than one galaxy concentration within the
measurement aperture (0.50 h$^{-1}$ Mpc) will have multiple density
peaks. Choosing the highest density peak for richness calculation --
{\it i.e.}, counting galaxies within 0.50 h$^{-1}$ Mpc from this new
center -- may not be appropriate for tracing the galaxy
distribution. If the cluster is defined by a single peak, then the
richness estimate should not be affected.

Beyond considering the effects mentioned above, we investigate which
metric aperture is optimal for computing the richness. In
\citet{gal03} we adopted a 1.00 h$^{-1}$ Mpc radius, which may not be
the best choice for the current purposes. Like \citet{pop04} we used the
scaling relations between optical and X-ray properties to guide the
choice of the optimal radius. We investigated how the scatter of these 
relations is affected by the different radii used for computing 
richness, examining the
L$_X-$N$_{gals}$ and T$_X-$N$_{gals}$ relations based on all clusters
common to BAX and DPOSS used in the substructure analysis
(see section 5).  Figure 2 shows the variation of the linear
correlation coefficient ($\rho$) and the orthogonal scatter ($\sigma$)
for these two relations with the aperture used to estimate
richness. We see a clear trend for minimum scatter and highest
correlation for an aperture of $\sim 0.50$ h$^{-1}$ Mpc. In Figure 3
we show the relation between the relative richness error
(N$_{gals-err}$/N$_{gals}$) and richness for two apertures. We see
that clusters with N$_{gals} < 10$ (within 0.50 h$^{-1}$ Mpc) have
errors comparable to their richnesses. The increase in the richness
error as we increase the radius from 0.50 h$^{-1}$ Mpc to 1.00
h$^{-1}$ Mpc is also striking. Based on these three plots and on Table
1 (where we show the scatter of the optical {\it vs} X-ray relations
for richness estimates based on different centroids and
redshifts) we re-compute the richness estimate (N$_{gals}$),
instead of using the values from Papers II and III.  We
adopt the original optical cluster position, the BAX redshift and an
aperture of $ 0.50$ h$^{-1}$ Mpc for each cluster, and use the BAX
cosmology. We also exclude clusters
with N$_{gals} < 10$ (unless otherwise stated) for the comparisons
shown in $\S$6. This step (re-computing richness) is crucial for
minimizing the scatter in the comparisons between optical and X-ray
properties ($\S$6 and Table 1). The impact of centroid, radius and redshift on
substructure measurement is discussed in $\S$5.

\begin{figure*}[!ht]
\includegraphics[totalheight=5.5in, scale=0.6, trim= 0 0.0in 0 0]{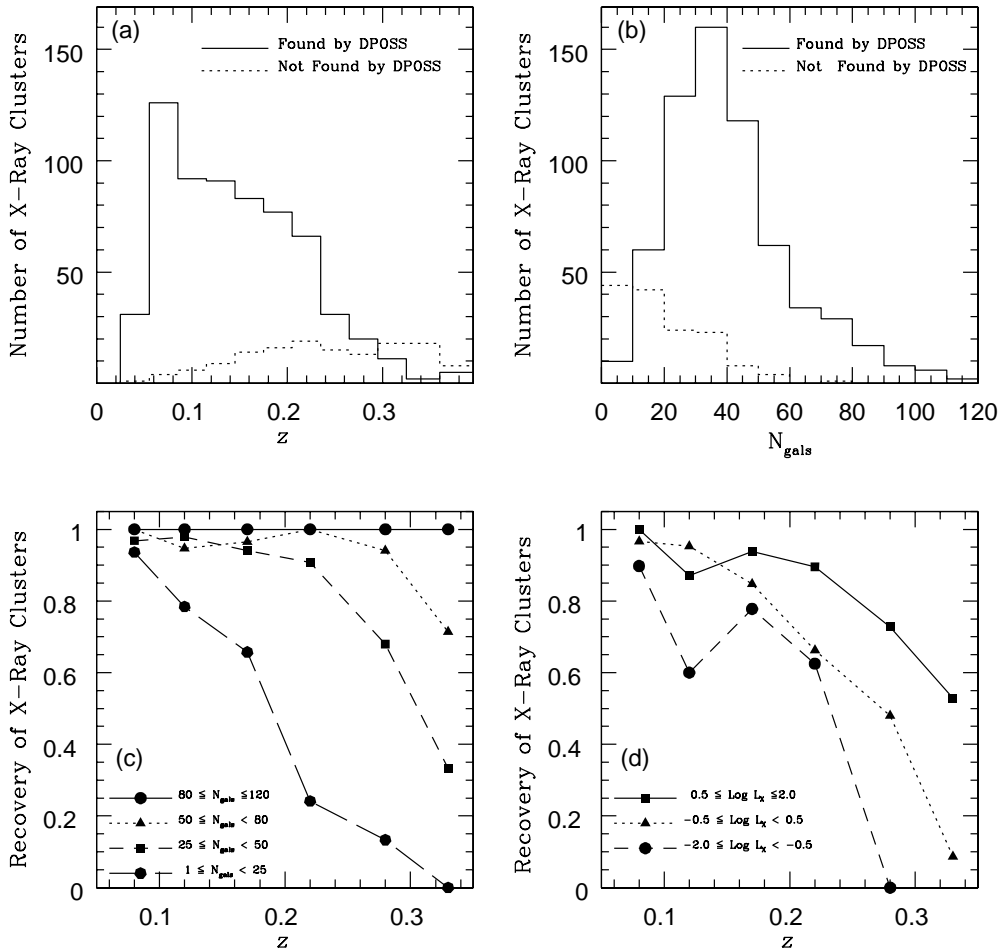}
\centering
\caption{The recovery rate of BAX clusters in NoSOCS. ({\it a}) redshift
distribution for common clusters and BAX clusters not found in DPOSS;
({\it b}) the richness distributions for common objects and BAX only sources;
({\it c}) recovery rate of X-ray clusters by 
NoSOCS, as a function of redshift, for different richness 
classes ($1 \le  N_{gals} < 25$; $25 \le N_{gals} < 50$; $50 \le N_{gals} 
< 80$; $80 \le N_{gals} \le 120$); ({\it d}) the same as in panel (c), but for
different X-ray luminosity ranges ($-2.0 \le Log L_X < -0.5$; 
$-0.5 \le Log L_X < 0.5$; $0.5 \le Log L_X \le 2.0$).
\label{fig2}}
\end{figure*}

\subsection{Comparison of the two catalogs}

We compare the BAX and NoSOCS catalogs using BAX
as the reference, searching for optical counterparts in NoSOCS. We
adopt a search radius of 1.50 h$^{-1}$ Mpc, keeping only the nearest
match to each of the 792 BAX clusters. The total number of recovered
systems is 638.  In Figure 4 we show the offset distribution in
h$^{-1}$ Mpc (solid line) and arcseconds (dotted line). The typical
X-ray$-$optical offset is $< 0.50$ h$^{-1}$ Mpc (or $< 400''$).  The
redshift and richness distributions of the common and missing clusters
are shown in Figure 5, as the top left and right panels, respectively.
The majority of BAX clusters not identified by NoSOCS are at higher
redshifts ($z \gsim 0.2$) and have low richnesses (N$_{gals} < 40$)
systems. For clusters not found by NoSOCS, we adopt the X-ray
coordinates (given in BAX) to compute their richnesses.

The overall recovery rate of BAX clusters by NoSOCS is $81\%$.
Considering the redshift and richness (or L$_X$) ranges sampled this
is an encouraging result. It is also important to keep in mind that
the BAX catalog is a compilation of all X-ray galaxy clusters found in
the literature and not a complete list to a given flux limit.  Instead
of trying to explain each missing cluster on a case-by-case basis, we
investigate their overall properties (richness and L$_X$) as a
function of redshift. This approach also enables us to directly
compare our results to the selection functions (SF) estimated by
\citet{gal03, gal06}.  The recovery rate of X-ray emitting clusters by
NoSOCS is shown in the lower panels of Figure 5. In the left panel we
show the detection rate for different richness classes, while in the
right panel the relation is shown for different X-ray luminosities.
Comparison of the left panel of this figure to Figure 6 from Paper II
suggests that the earlier simulations performed to evaluate the SFs
tend to {\it underestimate} the successful detection rate. However,
the BAX list used here is not a complete catalog and includes clusters
that were first found in the optical and only later observed at X-ray
wavelengths. Even considering these two points the results shown in
the lower panels of Figure 5 are quite impressive, demonstrating a
recovery rate of approximately 90$\%$ of all X-ray luminous (L$_X
\gsim 3.2$ $10^{44}$ $ergs$ $s^{-1}$) galaxy clusters at any redshift
out to $z \sim 0.2$. In terms of richness, we recover all X-ray
clusters cataloged in BAX that have N$_{gals} \gsim 80$ out to $z \sim
0.3$, and more than 90$\%$ of those with N$_{gals} \gsim 25$ to $z
\sim 0.2$.

\section{Optical Substructure}

We have used photometric data alone to obtain an estimate of the
fraction of clusters with evidence for substructure, based on our
sample of 10190 clusters at $0.07 \lsim z \lsim 0.21$. In order to
evaluate substructure we apply four two-dimensional (spatial)
statistical tests \citep{pin96}. It is well known that any
substructure test statistic has little meaning if not properly
normalized, which can be achieved by comparing the results for the
input data to those for substructure-free samples (the {\it null
hypothesis}). For the 2D tests employed in this work the null
hypothesis is given by an azimuthally symmetric, smooth distribution
of points, with surface density decreasing as a function of radius. We
generate such data via azimuthal randomization, where the distance of
each galaxy to the cluster center is maintained and a new azimuth is
randomly assigned. The main advantage of this technique is the exact
replication of the radial profile of the cluster. The four tests used
here each provide a significance level, given as the probability that
the observed distribution is drawn randomly from one free
of substructure.

The significance level is determined through Monte Carlo
simulations. For each input data file we generate $N=500$ simulated
data sets by azimuthal randomization. We then calculate the number of
Monte Carlo simulations which show more substructure than the real
data. Finally, this number is divided by the number of Monte Carlo
simulations. For most of the analysis in this paper we set our
significance threshold at 5$\%$, meaning that only 25 of the 500
simulated datasets can have substructure statistics higher than the
observations to consider a substructure estimate significant. This
choice of threshold is explained in $\S$5.5.

In the next four subsections we give a brief description of each
substructure estimator: the {\it angular separation test} (AST), the
{\it Fourier elongation test} (FE), the {\it Lee statistic} (Lee 2D)
and the {\it symmetry test} ($\beta$). Detailed descriptions of all
four tests are provided by \citet{pin96}.

\subsection{The Angular Separation Test}

The angular separation test was developed by \citet{wes88} and
considers the galaxy distribution in the cluster only in terms of
angular coordinates around the cluster center. In this case, the
presence of substructure is detected as an excess number of small
angular separations between galaxy pairs relative to the expected
number in a spherically symmetric, substructure-free distribution.
 The harmonic
mean angular separation is given by
$$\theta_{hm} = \biggl[2/N(N-1)
\sum_{i>j}\theta_{ij}^{-1}\biggr]^{-1}, \eqno(1)$$ where the sum is
performed over all pairs. N is the total number of galaxies and
$\theta_{ij}$ is the angular separation between galaxies $i$ and
$j$. The test statistic is the ratio of $\theta_{hm}$ measured for the
cluster and the harmonic mean for a Poisson distribution with the same
number of galaxies, AST = $\theta_{hm}$/$\theta_P$. This ratio is near
unity for substructure-free systems, and less than 1.0 for clumpy
distributions. $\theta_P$ is obtained from Monte Carlo simulations
whose sets are generated using the azimuthal randomization described
above. The test also adopts a small angle filter, rejecting all
$\theta_{ij}$ that are less than 1$\%$ of the expected mean
inter-galaxy separation if the N galaxies were uniformly distributed
(0.01[2$\pi$/N]).  \citet{wes88} find that this test is too sensitive
to the shape of the cluster density profile to provide unambiguous
evidence for the presence of substructure.  The main disadvantage of
this test is the loss of information when going from two spatial
dimensions to a single angular dimension. However, as pointed out by
\citet{pin96}, AST may be a more sensitive diagnostic for clumping in
multiple systems.

\subsection{The Symmetry Test}

The $\beta$ (or symmetry) test was also introduced by \citet{wes88} to
test for significant deviations from mirror symmetry about the cluster
center. This test assumes that a subcluster represents a local
asymmetry superposed on an otherwise symmetric distribution. For each
galaxy $i$, a local density estimate is obtained by the mean distance
to the $N^{1/2}$ nearest neighbors $d_i$ (\citet{wes88} used the five
nearest neighbors). The local density for a point $o$
diametrically opposite to galaxy $i$ is estimated in the same way
(estimating the mean distance to the $N^{1/2}$ nearest galaxies,
$d_o$).  For a symmetric galaxy distribution the values of $d_i$ and
$d_o$ should, on average, be approximately equal, but they will differ for
clumpy distributions.  The asymmetry for a given galaxy $i$ is
given by

$$\beta_i = log_{10}\biggl(\frac{d_o}{d_i}\biggr), \eqno(2)$$

The $\beta$ statistic is then defined by the average value
$<$$\beta_i$$>$ over all galaxies. For a symmetric distribution
$<\beta> \approx 0$, while values of $<$$\beta$$>$ greater than 0
indicate asymmetries. $<$$\beta$$>$ is a density-weighted average,
being more sensitive to the presence of substructure than if random
points were selected from the cluster. The
$\beta$ test is sensitive to deviations from mirror symmetry, but not
necessarily to ones from circular symmetry. In other words, the test does not
mistake an elongated smooth cluster for one containing substructure.

\subsection{The Fourier Elongation Test}

One of the characteristics of galaxy clusters is elongation, which can
be taken as a substructure indicator (although not a definitive one). 
In most cases, clusters with elongated 
galaxy distributions will have substructure. We use Fourier 
analysis \citep{rhe91} to estimate the elongation of galaxy clusters.
In this method we assume the azimuthal galaxy distribution ($N(\phi)$)
resembles the model 
$$N(\phi) = \biggl(\frac{N_0}{2\pi}\biggr)[1 + (N_1/N_0)cos(2\phi - 2\phi_0)], \eqno(4)$$
where $N_0$ is the number of cluster galaxies, $\phi_0$ 
is the cluster position angle and $N_1$ is the elongation amplitude. The 
statistic for this test is given by the elongation strength, defined by

$$FE = \frac{N_1}{\sqrt{2N_0}} = \frac{2(S^2 + C^2)^{1/2}}{\sqrt{2N_0}}, \eqno(5)$$

where

$$S = \sum N(\phi)sin(2\phi), C = \sum N(\phi)cos(2\phi). \eqno(6)$$

\citet{pin96} recommend a more strict criteria when using FE to detect
substructure. They advise using FE to reject the circular hypothesis
at the 1$\%$ level, with FE being greater than 2.5.  This is motivated
by the fact that elongation is also a signature of other formation
scenarios besides hierarchical mergers. Rather than adopt these
somewhat arbitrary criteria, we optimize this test the same way as the
others (section 5.5). The first criterion (1$\%$ rejection level) may
be too rigorous, while the second (selecting only clusters with FE $>$
2.5) makes little difference to the final results (Figure 9).

\subsection{The Lee Statistic}

The Lee statistic \citep{lee79} tests for bimodality in a given
distribution, which is perhaps the simplest type of substructure. In
our case, we test the hypothesis that the galaxies in a cluster lie in
two clumps. Since its introduction in 1979, the test has been used in
astronomical applications, {\eg} \citet{fit87} and
\citet{rhe91b}. The algorithm works as follows: (i) the data are
projected onto a line, which makes an angle $\phi$ with a second line,
assumed to be of constant declination. (ii) the first line is then
rotated by small increments for $0^\circ \le \phi \le
180^\circ$. (iii) For each orientation all points are projected onto
the second line, so that each point assumes a new coordinate, $x_j$, along
that line. (iv) A search for the best partition into two clumps
(named ``left'' and ``right'') is performed. The quantities $\sigma_l$,
$\sigma_r$ and $\sigma_T$ are calculated for the left, right and total
samples, respectively, for the $N-1$
partitions. The $\sigma$ are given by $\sigma = \sum(x_j - \mu)^2$,
with $\mu$ given by $\sum x_j/n$. The quantity $L$ is then defined by

$$ L = max_{partitions}\biggl(\frac{\sigma_T}{\sigma_l + \sigma_r} - 1
\biggr).  \eqno(7)$$ 

\noindent L varies with $\phi$, becoming large when the
projection axis connects two distinct clumps in the original
dataset. Following \citet{fit88} we adopt $L_{rat} = L_{max}/L_{min} =
maxL(\phi)/minL(\phi)$ as this test's statistic.  As in \citet{pin96}
we use $6^\circ$ increments in $\phi$. As pointed out by \citet{pin96},
the Lee 2D test is a conservative test because it does not detect
elongations as substructure and also loses sensitivity in the presence
of more than two subclusters (or if there are two clumps, but their
sizes are dissimilar).

\subsection{Summary of Substructure Tests}

The codes used for the substructure analysis are those of
\citet{pin96}, who evaluated the performance of 31 statistical tests
(including the four two-dimensional tests described above) on $N-$body
simulations of galaxy cluster mergers. For the two-dimensional tests
they recommend three diagnostics for substructure studies: FE, Lee 2D
and $\beta$. Based on the simulations, they find that the $\beta$ test
is the most sensitive, while the AST is very insensitive as it
requires a distinct compact subcluster to detect substructure. The FE
test is very sensitive to elongations, but they advise setting a
threshold for the FE statistic. The Lee 2D test requires the presence
of a very distinct subcluster, and is insensitive to more than two
clumps (or two clumps with very different sizes). \citet{pin96} point
out that Lee 2D is sensitive only to genuine two-dimensional
substructure, while FE and $\beta$ are able to find substructure in a
cluster which is not necessarily a merger. These conclusions are based
on their extensive simulations, and we do not have sufficiently
detailed data to independently test their assertions. We describe
below how we optimize the tests and the results when they are applied
to the DPOSS data.

\citet{pin96} conclude with a note saying that, when galaxies have no
redshifts, substructure could be erroneously detected due to the
superposition of foreground and background groups. While this is
certainly true, the large amount of two-dimensional data available
from the most recent sky surveys will suppress this effect.  The
acquisition of deep X-ray data or galaxy velocities for a large sample
of moderately redshift galaxy clusters is very expensive,
observationally speaking. Thus 2D data are also useful for the
selection of subsamples for future surveys \citep{kri97}.
\citet{kol01} used an alternate approach when computing the
fraction of optical substructure estimates that are due to background
contamination, searching for substructure using
2D optical and X-ray imaging data for 22 clusters.  They find good
agreement between optical and X-ray results, with only $\sim 22\%$ of
the clusters yielding inconsistent results (namely, substructure seen in the
optical, but not in the X-ray). They argue that this difference is
probably due to projection effects in the optical data. However, it is 
important to point out that some recent studies have found that 
projection can also affect the morphology inferred from X-ray data 
\citep{lop04, rin06}. In our case, we try to minimize this problem by 
using only galaxies selected in a given magnitude range (an issue 
that is not very well addressed in previous works). We have also 
optimized our methods to minimize the scatter when comparing optical 
and X-ray properties ({\eg} T$_X-$N$_{gals}$, described below).

\subsection{Dependence on Input Data}

The substructure results we show below are based on 10190 NoSOCS
clusters at $0.07 \lsim$ $z \lsim 0.21$, which is the largest sample
used for this type of study to date. The rationale for using only
clusters at these redshifts is detailed below. For the optimization of
the tests we have used the subsample of 445 clusters with a
counterpart in BAX. We first need to quantify the sensitivity of the
cited tests to the choices of centroid, redshift, magnitude range,
maximum radius and significance level by evaluating substructure for
different ranges of these parameters. In Figure 6 we show the effects
of the adopted center and redshift. We test four centers, the original
cluster position (O), the BAX coordinate (B), the luminosity weighted
center (L) and the position of the maximum density within the cluster
(P). The six panels in this figure are analogous to those in Figure 1,
now showing the residuals in the $\beta$ test. Additionally, we show
in the upper right of each panel the value of IQR for the residuals of
the significance of the $\beta$ test (not plotted; these are labeled
Q$_{\sigma2}$). The largest differences
are found when comparing the original cluster position to the BAX
coordinate (or the density peak position) and when adopting different
redshifts (BAX or DPOSS). The centroid choice affects the general
galaxy distribution, which may not be well indicated by the X-ray
center or maximum density peak (see the comments regarding Figure 1). The
redshift affects both the sampling of the cluster luminosity function
and the apparent radius used for selecting galaxies for the
substructure tests.

Based on this figure and on Table 1 we evaluate substructure (for the
entire sample of 10190 clusters) using the original cluster position,
instead of the luminosity weighted center or the density peak
coordinate.  For consistency, when using substructure to exclude clusters in
comparisons of optical and X-ray properties ($\S6$) we also adopt the
original cluster position and the BAX redshifts (see the richness
discussion). As shown in Table 1 this gives the minimum scatter for
the optical$-$X-ray relations based on the substructure results.

\begin{figure*}[!ht]
\centering
\includegraphics[totalheight=5.0in, scale=0.6, trim= 0 1.0in 0 0]{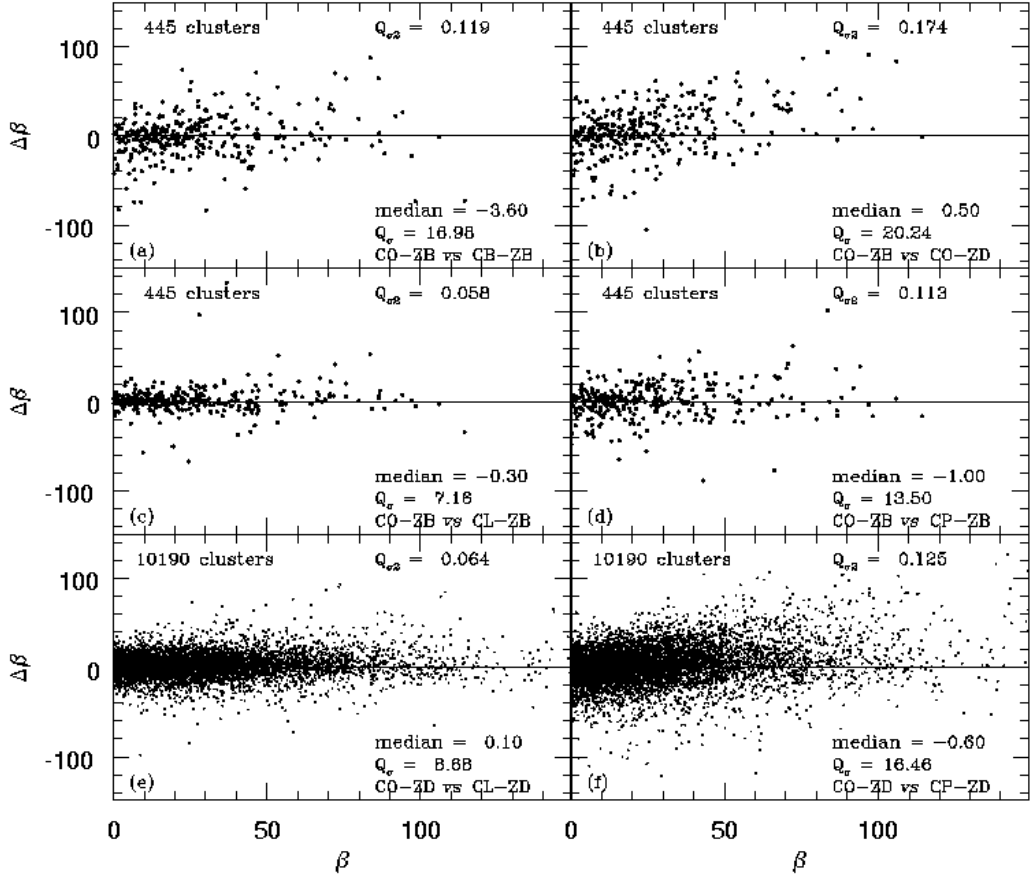}
\caption{The $\beta$ test residual for different cluster centroids 
and adopted redshifts. The comparison shown in each panel is analogous 
to those in Figure 1. The IQR value of the $\beta$ test residual
is labeled Q$_{\sigma}$. Additionally, we show in the upper right of 
each panel the values of IQR (named Q$_{\sigma2}$) for the residuals
of the significance of the $\beta$ test.
\label{fig6_02}}
\end{figure*}

In Figure 7 we show the dependence of the substructure results on the
magnitude range (lower panel) and maximum radius (middle panel) used
when selecting projected cluster galaxies.  Due to the large
computational cost of running these tests for the entire sample of
NoSOCS clusters, we show the dependence on the magnitude range and
maximum radius only for the subset common to BAX. Obviously, the
ability to detect substructure decreases when considering only the
bright central galaxies. Conversely, at fainter flux
limits and larger radii, the 2D data presents problems due to
background and foreground contamination.  We first attempted to
estimate substructure using all galaxies within the survey magnitude limits
($15.0 \le r \le 19.5$), but this leads to
misleading results due to different sampling of cluster galaxies for
clusters at different redshifts. We thus investigated the dependence
of the number of clusters with substructure on the magnitude range
used. We tested 6 ranges as indicated in the lower panel of Figure 7,
using only 100 (out of 638) clusters that span the full
magnitude range sampled ($m^*_r-2 \le m_r \le m^*_r+2$, corresponding to
redshifts of $0.11 \lsim z \lsim 0.14$).

We conclude that the high percentage of clusters with substructure
when using faint galaxies ($m^*_r+2$) may be an artifact due to
background contamination. We do not require the brightest galaxies
($m^*_r-2$) because that would further reduce our sample, eliminating
lower redshift clusters. As shown in Figure 7, the results for the
ranges $m^*_r-2 \le m_r \le m^*_r+1$ or $m^*_r-1 \le m_r \le m^*_r+1$ are
similar, having little effect on our conclusions. For all substructure
analyses we therefore use the latter range, which is
fully sampled by our data at redshifts $0.07 \lsim z \lsim 0.21$,
where we find 445 of the 638 common BAX and NoSOCS clusters (430 of
the 618 with estimates of L$_X$ and N$_{gals}$) and 53 of 101 with
estimates of T$_X$. From the 16546 clusters in NoSOCS there are 10190
in this redshift range.

From the middle panel of Figure 7 we see that out to 
R $= 1.00$ $h^{-1}$ Mpc the substructure tests behave similarly, 
except for AST, which is recognized by \citet{pin96} as 
being very insensitive. In the upper panel of Figure 7 we show 
the dependence of the number of clusters with 
substructure on the significance level (S.L.) required to reject 
the null hypothesis. Obviously, as we relax the threshold
to consider a substructure estimate meaningful, the number of selected clusters
increases. Our choice of significance level is explained below.

Any optical study relying on purely projected galaxy
positions is affected by superpositions of loose groups. One could 
assess projection effects in substructure estimates by
using X-ray data \citep{kol01} or including velocity dispersion
information. Given the large number of clusters in our sample, this
is not practical. Since our goal is to compare optical and X-ray
cluster properties ($\S$6) we optimize the parameters (magnitude
range, radius and S.L.) for the substructure tests by minimizing the
scatter in these scaling relations.  This could, in principle, lead to similar
results as those obtained by \citet{kol01}. This strategy is also
complementary to the analysis in Figure 7, which gives no objective
criteria to choose the optimal parameters.

This approach is similar to the one show in Figure 2. Before, we wanted to
find the optimal radius to compute richness (the one that minimizes the 
scatter of the scaling relations). Now we want to find the optimal radius,
magnitude range and S.L. to estimate substructure. Note that the optimal 
radius for substructure measurement need not be equal to the one used 
for richness, as we are measuring different cluster properties. 

\begin{figure}[!ht]
\centering
\includegraphics[totalheight=5.0in, scale=0.6, trim= 1.7in 0.4in 0 0]{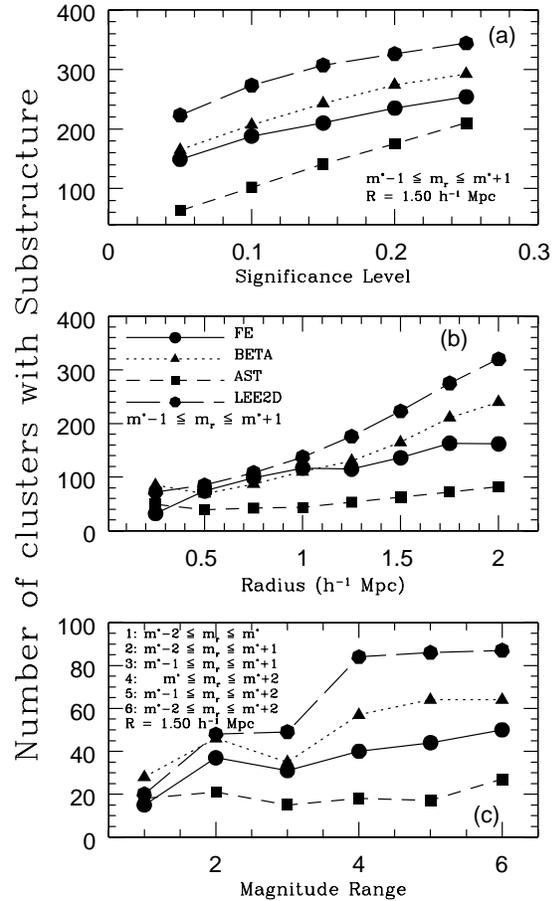}
\caption{Substructure results for the four statistical tests as a function
of the magnitude range (identified by an index from 1 to 6) sampled for each cluster (lower panel);
the radius (middle panel); the significance level required for 
each test (upper panel).\label{fig7_01}}
\end{figure}

In the upper panels of Figure 8 we examine how the orthogonal scatter
of three correlations of cluster parameters (L$_X-$N$_{gals}$,
T$_X-$N$_{gals}$ and T$_X-$L$_X$) can be minimized by excluding
clusters with signs of substructure. In the left panel, we show the
dependence of the scatter on the radius used for selecting cluster
galaxies, while on the right we show the variation with the
significance level employed. The results are shown only for the
$\beta$ test. As each test has a different dependence on these
parameters, their optimal values will also differ, implying that two
tests could yield the same percentage of clusters with substructure
through the use of very different apertures and especially
significance levels.  After many trials for all four tests we find
that the $\beta$ test is most appropriate (for our data) for detecting
substructure. It gives results that are comparable to the literature
and is considered the most sensitive among the four tests \citep{pin96}. 
Based on Figure 8, we use a radius of 1.50 $h^{-1}$ Mpc and a significance
level of 5$\%$. A similar analysis for the magnitude range suggests
that our original choice ($m^*_r-1 \le m_r \le m^*_r+1$) is
appropriate. 

It is interesting that only the T$_X-$N$_{gals}$ relation
is sensitive to the choices of radius and S.L.  As noted in Section 6,
the scatter in the L$_X-$N$_{gals}$ relation is much higher than for
T$_X-$N$_{gals}$. The exclusion of clusters based on substructure is
not sufficient to improve the L$_X-$N$_{gals}$ relation. In the lower
panels of Figure 8 we show similar plots, but for the four 
substructure tests and only for the T$_X-$N$_{gals}$ relation, using 
our final settings of R = 1.50 $h^{-1}$ Mpc and S.L. = 5$\%$.  From these 
two panels it is clear that only the $\beta$ test is sufficiently sensitive.

The distribution of substructure results, for the four statistical
tests, is shown in Figure 9. The solid lines represent the
distribution of all clusters, while the thick dotted lines show the
subset with substructure. A cluster is considered to have substructure
if the null hypothesis is rejected at the 5$\%$ significance level.
This means that at most 5$\%$ of the Monte Carlo datasets yield a
statistic more extreme than the real cluster.

\begin{figure*}[!ht]
\centering
\includegraphics[totalheight=5.5in, scale=0.6]{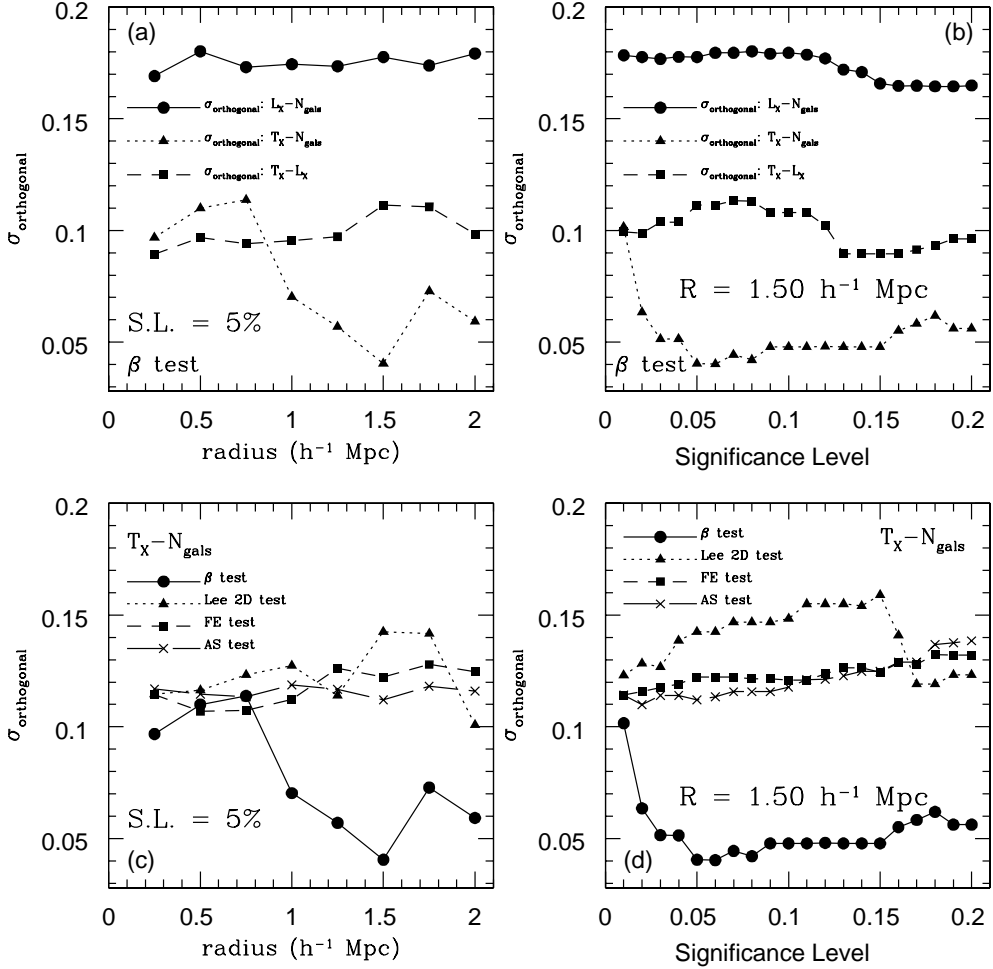}
\caption{The dependence of the orthogonal scatter on the radius
(left) and significance level (right) used to evaluate
substructure. In the upper panels we show this dependence for three
relations for the $\beta$ test only. The relations used are
L$_X-$N$_{gals}$, T$_X-$N$_{gals}$ and T$_X-$L$_X$. In the lower
panels the dependence is shown only for the T$_X-$N$_{gals}$ relation
for all four tests. \label{fig9}}
\end{figure*}

\begin{figure*}[!ht]
\centering
\includegraphics[totalheight=5.0in, scale=0.6, trim= 0 1.2in 0 0]{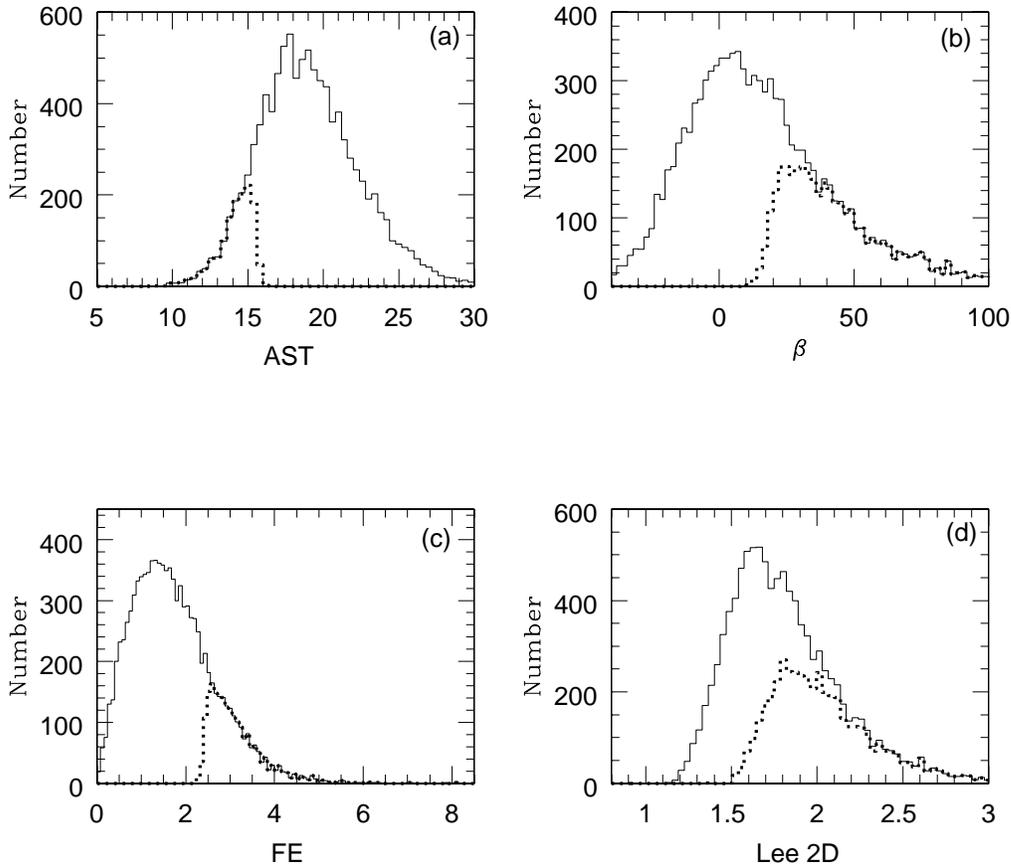}
\caption{Distribution of the substructure results for the 
four statistical tests: ({\it a}) the angular separation test (AST);
({\it b}) the $\beta$ test (to improve the visualization the $\beta$ 
values are multiplied by 1000); ({\it c}) Fourier Elongation (FE) and 
({\it d}) Lee 2D results.
The solid lines represent the 
distribution of all clusters, while the thick dotted lines show the results
for the subset with substructure based on the Monte Carlo simulation, rejecting the null hypothesis
at the 5$\%$ significance level. Note that AST is the only test for
which lower values represent a more extreme statistic. For the FE test
we show the distribution of all clusters considered to have substructure at 
the 5$\%$ significance level, without further excluding clusters with 
FE $>$ 2.5. As we can see the adoption of this additional criterion -- 
as suggested by \citet{pin96} -- excludes only a small additional number of clusters.
\label{fig7_02}}
\end{figure*}

We have also investigated the possibility that the substructure tests
depend on the cluster contrast, richness and redshift, with the
results shown in Figure 10. As mentioned in $\S$4.1, the contrast is
defined as the ratio between the number of galaxies in the cluster
region (within the same magnitude range sampled for the richness
calculation) and the background error $Q_{\sigma_{bkg}}$. There is no
clear sign that substructure detection is contrast-dependent. The
variation in the fraction of clusters showing substructure with the
cluster contrast is shown in lower panel, the redshift variation in
the middle panel, and the richness dependence in the upper panel. We
can not verify a clear trend with contrast, but see a mild increase
with richness and a very small decrease with redshift.  Since this
analysis is restricted to clusters at $0.07 \lsim z \lsim 0.21$, it is
difficult to investigate the evolution of substructure with look-back
time. We postpone the study of the evolution of substructure with
redshift to a future work, where we will use deeper photometry
(reaching higher redshifts) combined with X-ray imaging.

The final substructure results using the 10190 clusters at $0.07 \lsim
z \lsim 0.21$, for each test are as follows: AST finds that 13$\%$
clusters have substructure; the $\beta$ test selects 35$\%$; the FE
test finds 21$\%$; and the Lee 2D test finds 45$\%$, all at the 5$\%$
significance level. As we can see, AST selects by far the fewest
clusters with substructure, as expected due to its lack of
sensitivity. The rates found by the $\beta$ and Lee 2D tests are in
good agreement with the results of \citet{kol01}, who found at least
$\sim 45\%$ of their clusters with strong indications of
substructure. However, their sample was much smaller (only 22
clusters), and they employed other tests, using optical and X-ray data
to confirm the substructure estimates.  Other literature results
indicate that the percentage of clusters showing substructure varies
from 30$\%$ to 80$\%$. This discrepancy is generally believed to be
associated to the method employed for each analysis \citep{kol01}. We
stress that the variation in these results is likely also due to the
cluster centroid, area, magnitude range and significance level adopted
when applying a given test. For instance, had we used 1$\%$ or 10$\%$
as our significance level for the $\beta$ test, we would find 19$\%$
and 46$\%$ of clusters with substructure, respectively. Thus, not only
it is fundamentally important to compare similar datasets, the tests
employed must be consistent. The same clearly holds true for
comparisons with cosmological simulations.

\subsection{Correlation Between Substructure and Ellipticity}

Ellipticity is generally considered to be a sign of substructure, a
result that has been found by previous authors \citep{jon99,
kol01}. We thus decided to investigate if the substructure tests we
employ in this work are correlated to ellipticity. For this purpose we
used only the 445 common BAX--NoSOCS clusters at $0.07 \lsim z \lsim
0.21$. In order to compare our results to the ones from \citet{kol01},
we have considered substructure measures within 0.75 $h^{-1}$ Mpc, an
aperture close to the maximum radius employed by these authors (0.60
$h^{-1}$ Mpc). We have also binned the substructure and ellipticity 
results into substructure bins. For the four tests used in this work 
we fit a linear correlation of the type $\epsilon$ = A + Bsub$_{test}$, where
$\epsilon$ is the ellipticity value, sub$_{test}$ is the result for a
given substructure test, and A and B are the intercept and slope
obtained by an orthogonal regression method (see $\S$6.2). The relations
obtained for all the four substructure tests are shown on Figure 11 and
summarized in Table 2. We note that two of the methods (FE and Lee2D) show
a clear dependence between ellipticity and substructure. These results
are in agreement to the ones found by \citet{kol01}, although we find slopes
shallower than what they found. However, it is important to keep 
in mind that they employed a different substructure test and selected 
galaxies from a smaller aperture and different magnitude range.

\begin{figure}[!ht]
\centering
\includegraphics[totalheight=5.0in, scale=0.6, trim= 1.7in 0.6in 0 0]{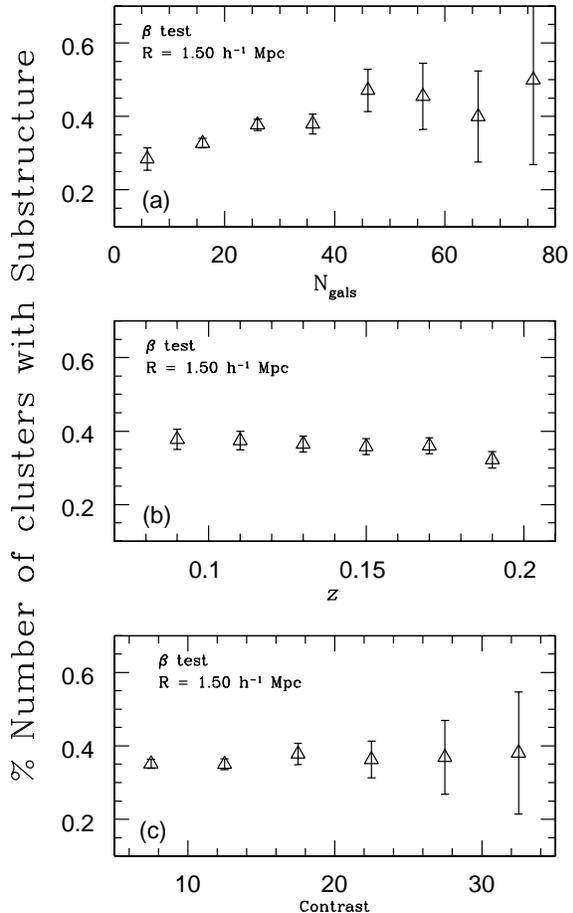}
\caption{Substructure results for the $\beta$ test as a function
of cluster contrast (lower panel), redshift (middle panel) and
richness (upper panel).
\label{fig8}}
\end{figure}

As can be seen from Figure 9, the $\beta$ test can take on negative values.
We found that these are anti-correlated to $\epsilon$, while the positive
values are correlated. We have thus considered only the positive values of
$\beta$ for the results shown in Table 2 and Figure 11. From inspection 
of this table we see that AST is the only test that  yields a slope
consistent to zero. This result is consistent with the aforementioned 
lower sensitivity of this test (Figure 7).

\begin{figure*}[!ht]
\centering
\includegraphics[totalheight=5.0in, scale=0.6, trim= 0 1.2in 0 0]{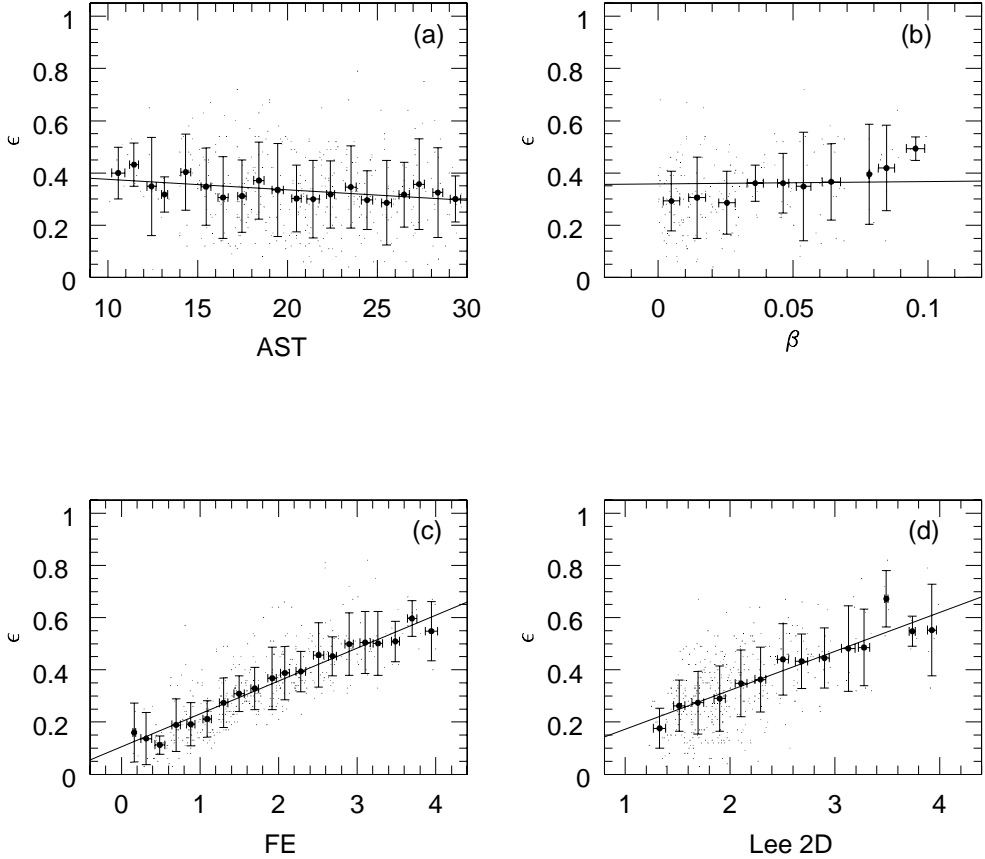}
\caption{Correlation between ellipticity and substructure for the 
four statistical tests: ({\it a}) the angular separation test (AST);
({\it b}) the $\beta$ test; ({\it c}) Fourier Elongation (FE) and 
({\it d}) Lee 2D results.
\label{fig8}}
\end{figure*}

\begin{deluxetable}{ccccc}
\tabletypesize{\footnotesize}
\tablecolumns{5}
\tablewidth{0pc}
\tablecaption{Correlation between ellipticity and substructure.}
\tablehead{
\colhead{Subst. Test} & \colhead{Linear Corr. Coef.} & \colhead{Intercept} &
\colhead{Slope}
}
\startdata
$\beta$ & 0.92 & 0.358 $\pm$ 0.017 & 0.092 $\pm$ 0.052 \\
Lee2D & 0.94 & 0.023 $\pm$ 0.037 & 0.149 $\pm$ 0.017 \\ 
FE & 0.98 & 0.105 $\pm$ 0.014 & 0.126 $\pm$ 0.007 \\ 
AST & -0.59 & 0.415 $\pm$ 0.025 & -0.004 $\pm$ 0.001 \\
\enddata
\end{deluxetable}

\section{Optical {\it vs.} X-ray Cluster Properties}

In this section we investigate the correlation between richness 
(N$_{gals}$) and X-ray properties (L$_X$ and T$_X$) of clusters common to
DPOSS and BAX. From the 638 common clusters, 620 have 
L$_X$ (of which 2 have no richness estimate), while T$_X$ is available for 101 clusters. As mentioned
before, for analyses involving substructure estimates
we restrict the study to $0.07 \lsim z \lsim 0.21$, reducing
the original samples from 618 to 430 clusters (with L$_X$ and richness
determined) and from 101 to 53 systems (with T$_X$).
Clusters with N$_{gals} < 10$ are also excluded when 
establishing these correlations
(see the discussion in $\S$4.1 and Figure 3) and a 
3$-\sigma$ clipping is also applied to exclude outliers.
We also discuss how the scatter in the established relations is
affected by the radius adopted for calculating richness, 
the redshift difference ($\Delta z$) between BAX and
DPOSS, positional offsets, the contrast of the cluster, and 
the presence of substructure.

\begin{figure*}[!ht]
\includegraphics[totalheight=4.0in, scale=0.6, trim= 0.2in 3.5in 0 0]{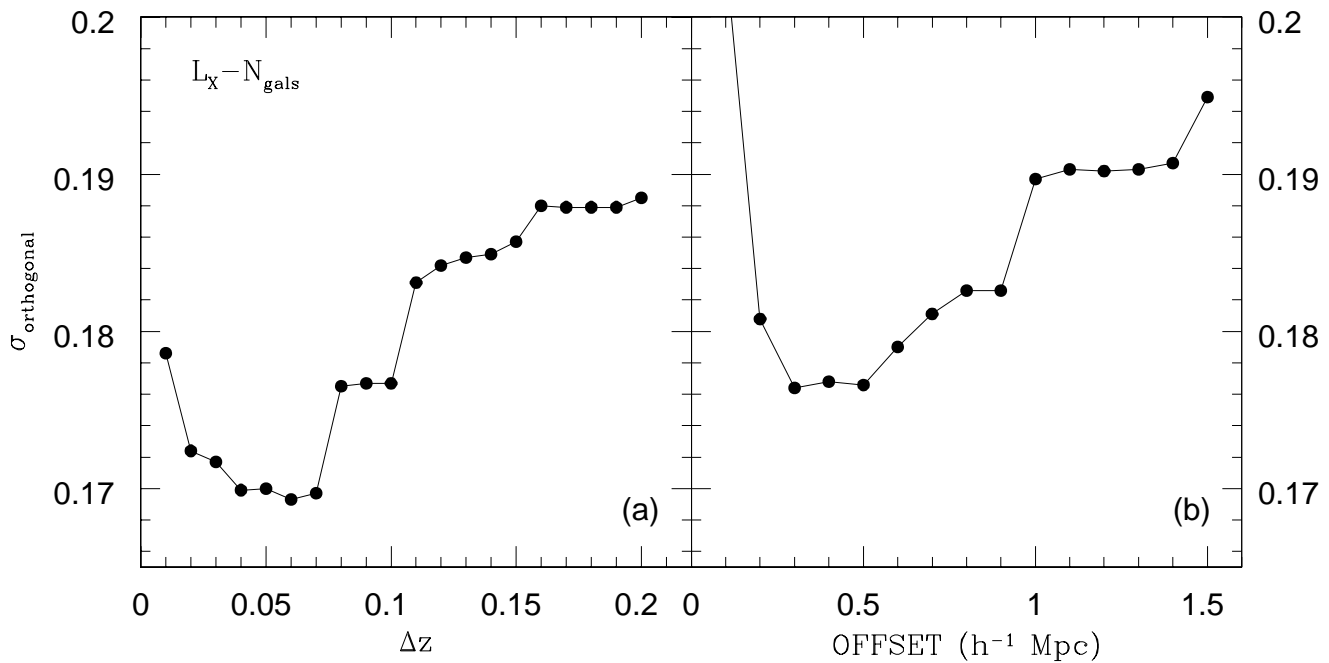}
\caption{Variation of the scatter of the L$_X-$N$_{gals}$ relation
as a function of the maximum absolute redshift difference between
the NoSOCS and BAX values (left panel) and the positional 
offset (right panel).
\label{fig5}}
\end{figure*}

\subsection{Dependence on Input Data}

We reiterate that the orthogonal scatter ($\sigma$) and the linear
correlation coefficient $\rho$ of the L$_X-$N$_{gals}$ and
T$_X-$N$_{gals}$ relations are extremely sensitive to the aperture
used for calculating richness (see Figure 2).  These results are in
good agreement with those of \citet{pop04}, who verified the
dependence of the scatter in the L$_X-$L$_{opt}$ relation on the
aperture used. When using very small apertures to estimate N$_{gals}$
the galaxy number count becomes very uncertain due to centroid
errors. For large radii the background contribution grows rapidly,
contaminating the richness estimates. As explained in $\S$4.1 we use a
0.50 $h^{-1}$ Mpc radius for estimating N$_{gals}$ for comparisons to
X-ray properties.

The effects of $\Delta z$ and positional offset on the scatter in the
L$_X-$N$_{gals}$ relation are shown in the left and right panels of
Figure 12, respectively.  As before, we adopt the BAX redshift to
calculate N$_{gals}$, since the richness is strongly
redshift-dependent (through the apparent radius and the magnitude
range). This should minimize the scatter in the L$_X-$N$_{gals}$
relation, and we expect this scatter to show no dependence on $\Delta
z$ if the redshift errors are random. The variation shown in the left
panel of Figure 12 is small ($\sim 0.020$) but real,  and is probably
due to the fact that the BAX redshifts are not homogeneous
measures. They are automatically retrieved from NED, and are most
often spectroscopic, but some are photometric estimates. Even the
spectroscopic redshifts may be based on only one or two galaxy
spectra. Thus, we conclude that even after adopting the BAX redshift,
we could further reduce the scatter in the L$_X-$N$_{gals}$ relation
by restricting the sample to clusters with small values of $\Delta z$
(at the cost of reducing the sample size). The improvement in the
least-square fit of the L$_X-$N$_{gals}$ relation for clusters with
small positional offset is expected. The dependence of
$\sigma_{orthogonal}$ is of the same order as that shown in the left
panel.

\begin{figure*}[!ht]
\centering
\includegraphics[totalheight=5.0in, scale=0.6, trim= 0 1.2in 0 0]{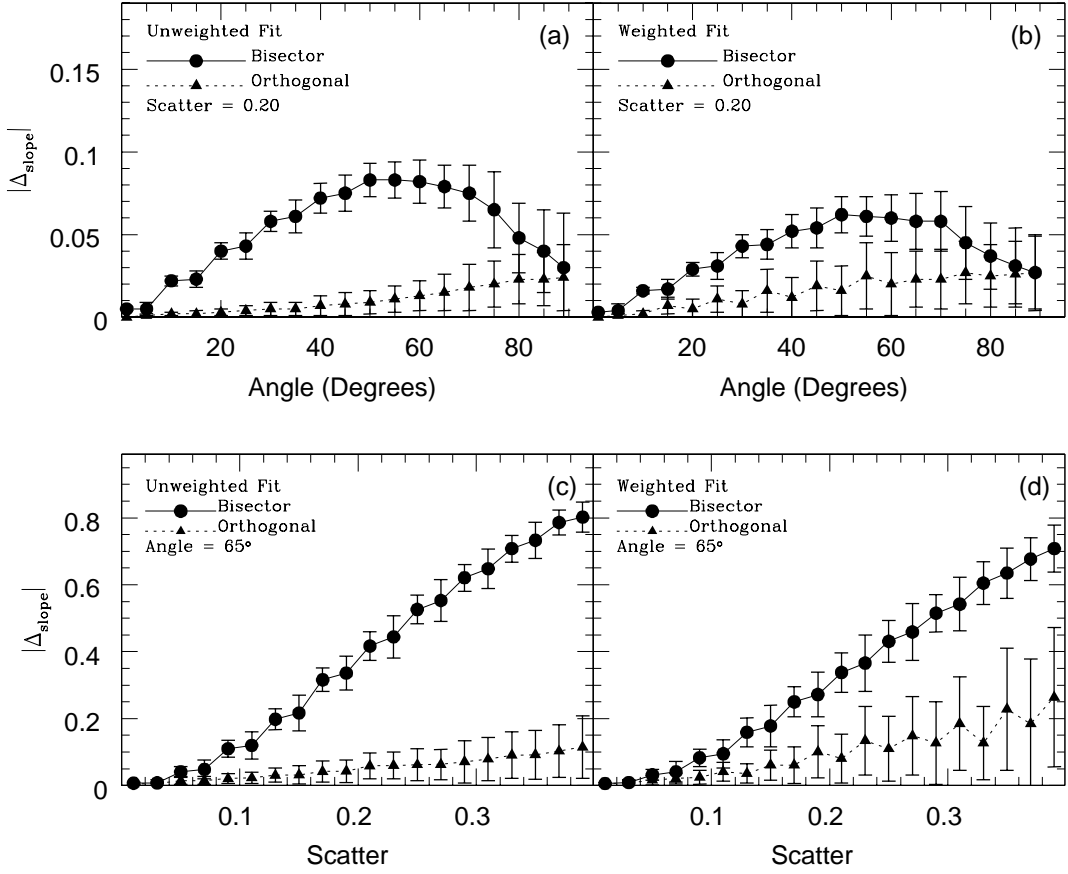}
\caption{Variation of $|\Delta_{slope}|$, the difference between the input
slope and the one recovered by the bisector and orthogonal solutions. In the
upper panels we show the variation with  input angle (at a fixed scatter of 0.20), while in the lower panels we show the variation with scatter 
(at a fixed angle of 65$^{\circ}$). The left panels show the results for the
unweighted fits and the right panels for the weighted fits. The bisector 
results are given by the solid lines, while the orthogonal ones are shown
by the dotted lines.
\label{fig11}}
\end{figure*}

We also investigate the dependence of the L$_X-$N$_{gals}$ with
the minimum and maximum redshifts used to select clusters for the
fit. The results are optimal for clusters at $0.15 \le z \le 0.30$,
with strong sensitivity to the lower redshift limit. This is because
most of the low richness systems are at low redshift, and these poor
clusters have the largest richness errors (Figure 3). Based on Figures
2 and 12 we use a radius of $ 0.50$ $h^{-1}$ Mpc for calculating
richness, but do not apply any cut with redshift or offset. All the
dependences of $\sigma_{orthogonal}$ on the redshift cutoff, $\Delta
z$ and the positional offset are much less pronounced than the
dependence on aperture. Thus, we do not restrict the sample based on
these weaker effects, which would significantly reduce the number of
clusters and make it difficult to quantify the most significant
effects. The impact of contrast and substructure are investigated
further below.

\subsection{Dependence on Fitting Method}

Before examining the connection between optical and X-ray parameters
we investigate which linear regression solution is more robust, using
the packages {\it slopes} and {\it bces-regress} from \citet{iso90}
and \citet{akr96}, respectively.  The first performs unweighted fits,
while the second considers errors as well as intrinsic
scatter. Because BAX provides no error estimate for L$_X$, we perform
an unweighted fit (using {\it slopes}) for the L$_X-$N$_{gals}$
relation, while {\it bces-regress} is used for T$_X-$N$_{gals}$. We
use two commonly used solutions for each package. The first is the
ordinary least-square (OLS) bisector solution, the line which bisects
the OLS(Y$|$X) (minimizing the residuals in Y) and OLS(X$|$Y)
(minimizing the residuals in X) lines.  The second solution is the
orthogonal regression line, which minimizes the perpendicular
distances \citep{iso90, fei92, akr96}.  We stress the comment of
\citet{iso90} concerning linear regression methods that the two
methods used here (and others available in the literature) find
regression coefficients that are theoretically different from each
other. In other words, they do not represent different estimates of
the same quantity. Even if the entire population were sampled, the
measured slopes would differ (only in special cases is a single
relation obtained, such as when $\rho = 1$).

In order to evaluate which of these methods is best suited for our
purposes, we generated artificial data sets with known scatter and
slopes. The relations resulting from the optical -- X-ray comparison
could be very steep and noisy, especially those involving L$_X$. Hence, we
created data sets of 1000 points for different angles (ranging from 1
to 89 degrees) and for 20 different scatters (from 0.01 to 0.40). We
then run {\it slopes} and {\it bces-regress} for the bisector and
orthogonal solutions, computing the difference between the input and
output slopes ($|\Delta_{slope}|$). This procedure is repeated 100
times. The results are shown in Figure 13, where the error bars for each
point represent the standard deviation for the 100 events. In the upper 
panels we show the variation of $|\Delta_{slope}|$ with 
angle (at a fixed scatter of
0.20), while the dependence with scatter is shown in the lower panels
(for a fixed angle of 65$^{\circ}$). The tests using unweighted and
weighted fits are shown in the left and right panels,
respectively. The bisector results are plotted as solid lines while
the orthogonal solutions are the dotted lines.  As with the real data,
a 3$-\sigma$ clipping is applied to exclude outliers. The orthogonal
minimization is clearly superior for recovering the input slopes,
especially when there is large scatter. If we do not eliminate
outliers, the results shown in the upper panels diverge for large
angles, but the bisector solution still shows larger deviations from
the original slopes. Given these results, we use the orthogonal
regression for all fits. We note that some results in the literature
\citep{mar98} have used the bisector, which may be misleading.

\subsection{The X-ray and Optical Relations}

\subsubsection{X-ray Luminosity and Temperature}

We begin our analysis by examining the L$_X-$T$_X$ relation (Figure
14) derived from the BAX data, to verify the reliability of this data
source for our purposes. We use 120 clusters with available
temperatures within the NoSOCS region.  As the solid thin line we show
the best fit using all 120 clusters (two are excluded as outliers). As
the thin dotted line we show the results obtained using only the 100
clusters with optical contrast $\ge 10.0$. As discussed below, the
scatters and slopes of the scaling relations depend on the contrast
cut applied to the cluster sample. Interestingly, the optical contrast
appears to provide a useful test for selecting clusters to affect the
L$_X-$T$_X$ relation. Literature results are shown by thick lines.
The solid and dotted lines show the results from \citet{mar98}, with
the dotted line for L$_{X,bol}$. We note that \citet{mar98} employed a
bisector method for fitting this relation. If we also use the bisector
slope, our results are considerably less steep (L$_X \propto
T{_X}^{2.71}$), showing better agreement with \citet{mar98}. However,
as discussed above, the bisector is not the optimal method for fitting
steep slopes especially in the presence of large scatter. The results
from \citet{dav93} are plotted as the dotted thick line.

\begin{figure*}[!ht]
\centering
\includegraphics[totalheight=5.0in, scale=0.6, trim= 0 0 0 0]{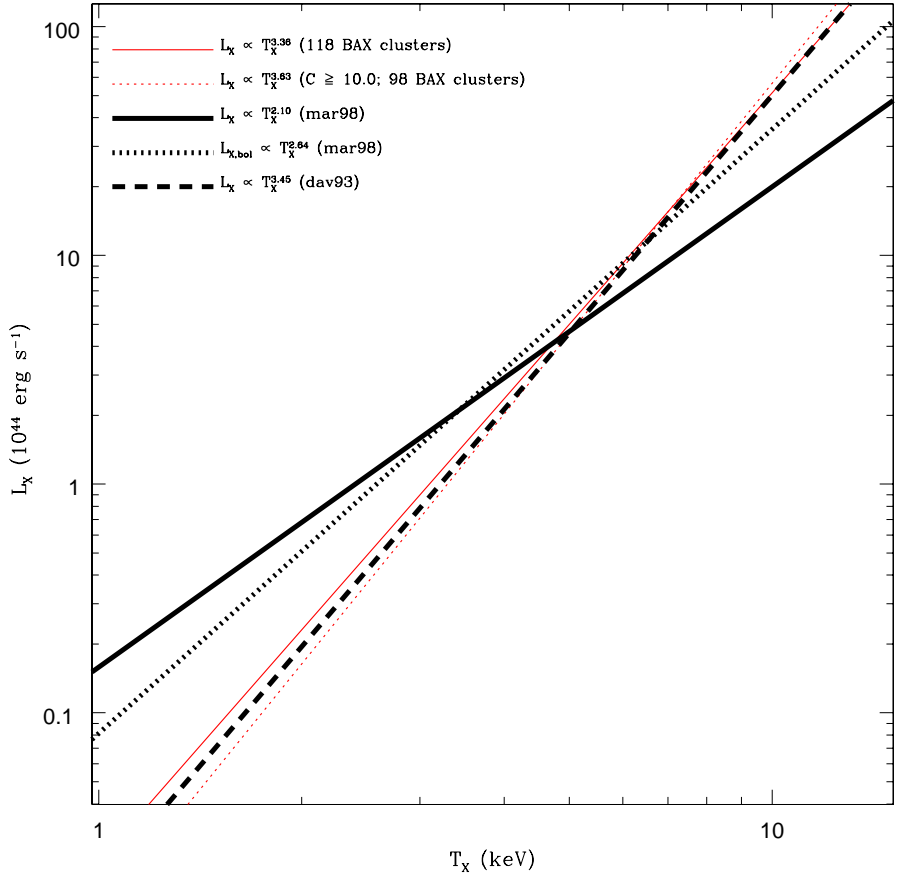}
\caption{The L$_X$ - T$_X$ relation obtained using all 120 BAX
clusters (with two excluded as outliers) in the NoSOCS region is shown
as a thin solid line. The sample is then restricted to high optical
contrast systems (C $\ge$ 10.0) with the result indicated by the thin
dotted line. Literature results are indicated by thick lines. The
relations obtained by \citet{mar98} are indicated by the solid and
dotted lines (the last one uses bolometric luminosities). The dashed
line indicates the results from \citet{dav93}.
\label{fig12}}
\end{figure*}

\subsubsection{X-ray Luminosity and Richness: Structure Segregation}

The L$_X-$N$_{gals}$ relation is shown in Figure 15, along with the
effects of the contrast and the presence of substructure in the
derived relation. The comparison of all clusters at $0.07 \lsim z
\lsim 0.21$ is exhibited in the lower left panel, while the high
contrast systems (contrast $\ge$ 12.0) are shown in the lower right.
The substructure--free clusters are exhibited in the upper left panel
while those with substructure in the upper right.  The fit shown by a
solid line in all panels is that obtained for the substructure--free
clusters. In panels (b), (c) and (d) we also show the 3$\sigma$
boundaries of this fit (as dotted lines) and the fit obtained with the
data shown in each panel (all clusters, high contrast systems and
clusters with substructure) with the dashed line.  In all panels we
indicate the number of clusters used in the fits (after excluding
outliers at a 3$-\sigma$ level), the linear correlation 
coefficient ($\rho$), the intercept
(A), slope (B) and the orthogonal scatter of the relation obtained for
that sample.

From this figure we see that the scatter changes only slightly when selecting only substructure--free clusters, while it is 
reduced by $\sim 12\%$ when selecting high contrast systems.
The combination of the two criteria (substructure and contrast) does not
improve the scatter compared to applying only the 
contrast cut. Interestingly, the slope is significantly higher for the
high contrast clusters. \citet{pop04} also found improvement in the relations
when excluding low contrast systems.
We conclude that the scatter in the L$_X$ -- N$_{gals}$ relation is
mainly due to the difficulty of obtaining accurate measures of
these parameters for low luminosity systems. If there is any difference
for clusters with or without substructure, it is washed out 
in these relations. Thus, when correlating L$_X$ to the optical 
parameters we do not detect any significant structure segregation.

\begin{figure*}[!ht]
\centering
\includegraphics[totalheight=5.5in, scale=0.6]{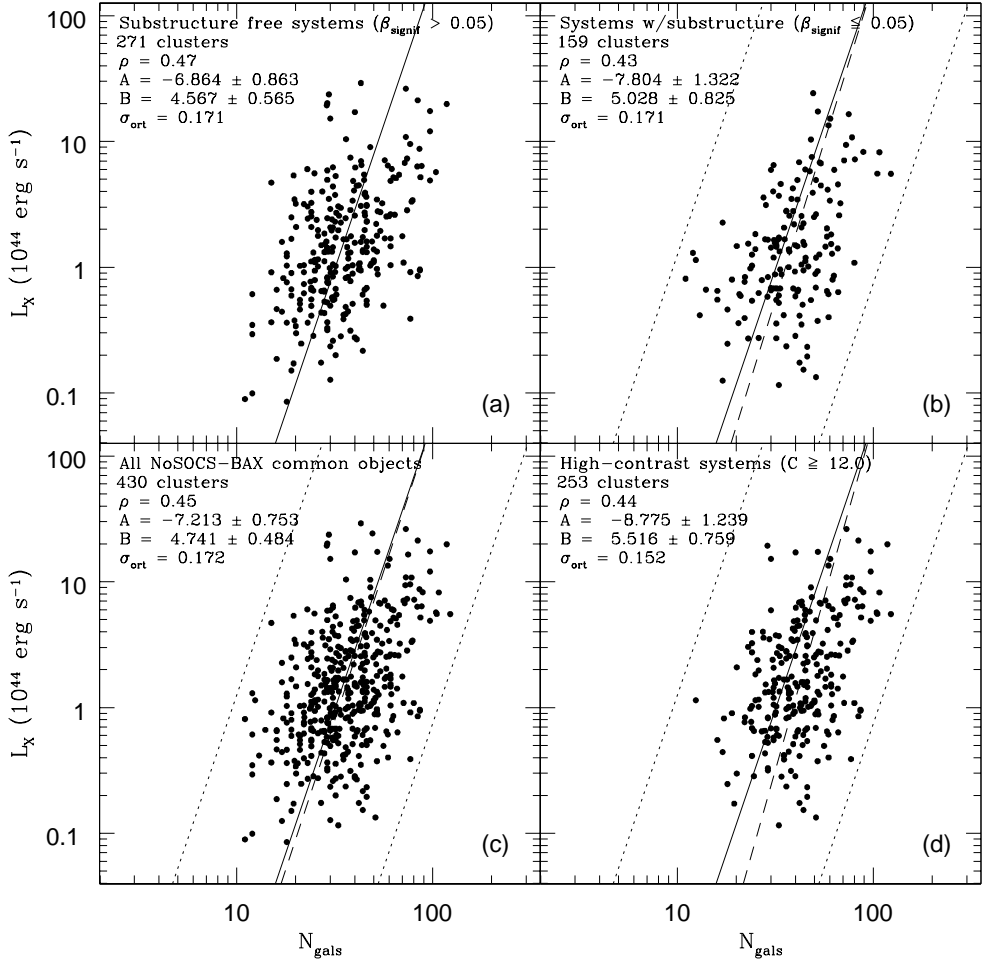}
\caption{The L$_X-$N$_{gals}$ relation is shown for: ({\it a}) all clusters
at $0.07 \lsim z \lsim 0.21$; ({\it b}) clusters with contrast $\ge 12.0$;
({\it c}) clusters without substructure (at the significance 
level of $5\%$) based on the $\beta$ test; ({\it d}) clusters with substructure according to the $\beta$ test. The fit shown in all 
panels (as a solid line) is the orthogonal solution obtained for the 
sample in the upper left panel. For 
three panels (a, b and d) the dotted lines show the 3$\sigma$ 
boundaries for this fit. For these 3 panels (a, b and d) the fit 
obtained with the data shown on each panel (all clusters, high
contrast systems and clusters with substructure) is shown by the 
dashed line.
\label{fig13_01}}
\end{figure*}

\subsubsection{X-ray Temperature and Richness: Structure Segregation}

The T$_X-$N$_{gals}$ relation is plotted in Figure 16 (with 3$-\sigma$ 
outliers shown as open symbols), in a manner analogous to the previous Figure.
In the lower left panel of Figure 16 we show the results for the 53
clusters at $0.07 \lsim z \lsim 0.21$ with T$_X$ and richness
measured. As before, in the lower right panel we show the results for
the high contrast systems (contrast $\ge$ 12.0). In the upper left the
sample includes only clusters found to have {\it no} substructure by
the $\beta$ test at the 5$\%$ S.L, while in the upper right we show
the complementary clusters. A few things stand out from this
figure. First, T$_X$ is clearly better correlated with the optical
parameters than L$_X$. Second, substructure has an effect on the
scatter of the T$_X-$N$_{gals}$ relation, which was not seen when
using L$_X$. When excluding clusters with substructure the scatter of
the T$_X-$N$_{gals}$ relation is reduced by $\sim 63\%$. Third, there
is no improvement when applying the contrast cut. This is mainly due
to the fact that T$_X$ is typically available only for the most
massive systems, so the analysis is restricted to a small range of
temperature (and also contrast).  Of the 430 clusters with L$_X$
measured, 177 ($\sim 41\%$) clusters are low contrast systems, while
they are only 14 ($\sim 26\%$) of the 53 with T$_X$ . Finally, when
combining the two criteria (substructure and contrast) the scatter is
not very different than if we employ only the substructure cut,
implying that the dominant factor in the scatter of this relation is
substructure.

The small number of clusters with available temperatures is certainly
a problem when establishing these relations and applying the above
cuts. However, \citet{smi05} have also recently found that the scatter
in the scaling relations between cluster mass, X-ray luminosity and
temperature is dominated by unrelaxed clusters. Their sample is
smaller than ours (and they are also limited to massive systems), but
serves to corroborate the current results. An opposite result (no
structure segregation in the scaling relations) is found by
\citet{oh06}, using both observational data and simulations. They
compare the deviations of clusters from the best fit relations
(obtained for all systems and not for the separate sets with or
without substructure) for different scaling relations and find no
evidence for higher scatter for clusters with more substructure.  The
substructure tests they employ are different than the ones used
here. Their method for dividing the sample into clusters with or
without substructure is also unclear (there is no cut at a given
significance level).  However, the most important differences between
our methodology and theirs are the fact that they investigate
substructure with an aperture that scales with the cluster mass
($R_{500}$ and $R_{2500}$), while we use a fixed metric ($ 1.50$
$h^{-1}$ Mpc radius), and that they rely on X-ray data, while we use
optical data. Clusters with more substructure are
probably dynamically younger and would thus be expected to deviate
from relations established for well relaxed systems. In future work,
we will investigate substructure using these alternative approaches \citep{oh06}
to see if our current results persist.

It comes as no surprise that the X-ray temperature correlates better
with optical properties than the X-ray luminosity. T$_X$ has been
shown to provide a much tighter correlation to cluster mass than X-ray
luminosity \citep{voi05}, as T$_X$ is closely related to the cluster's
potential well. What we find here is that the connection between
richness (which is a mass tracer) and T$_X$ has a smaller scatter than
if we used L$_X$. Similar results are found by \citet{yee03} and
\citet{pop05} who find tighter relations between their richness
estimates (B$_{gc}$ or L$_{opt}$) and T$_X$
than with L$_X$. We also note that samples constructed purely at one wavelength may not include the same types of objects \citep{lub04}.

\begin{figure*}[!ht]
\centering
\includegraphics[totalheight=5.5in, scale=0.6]{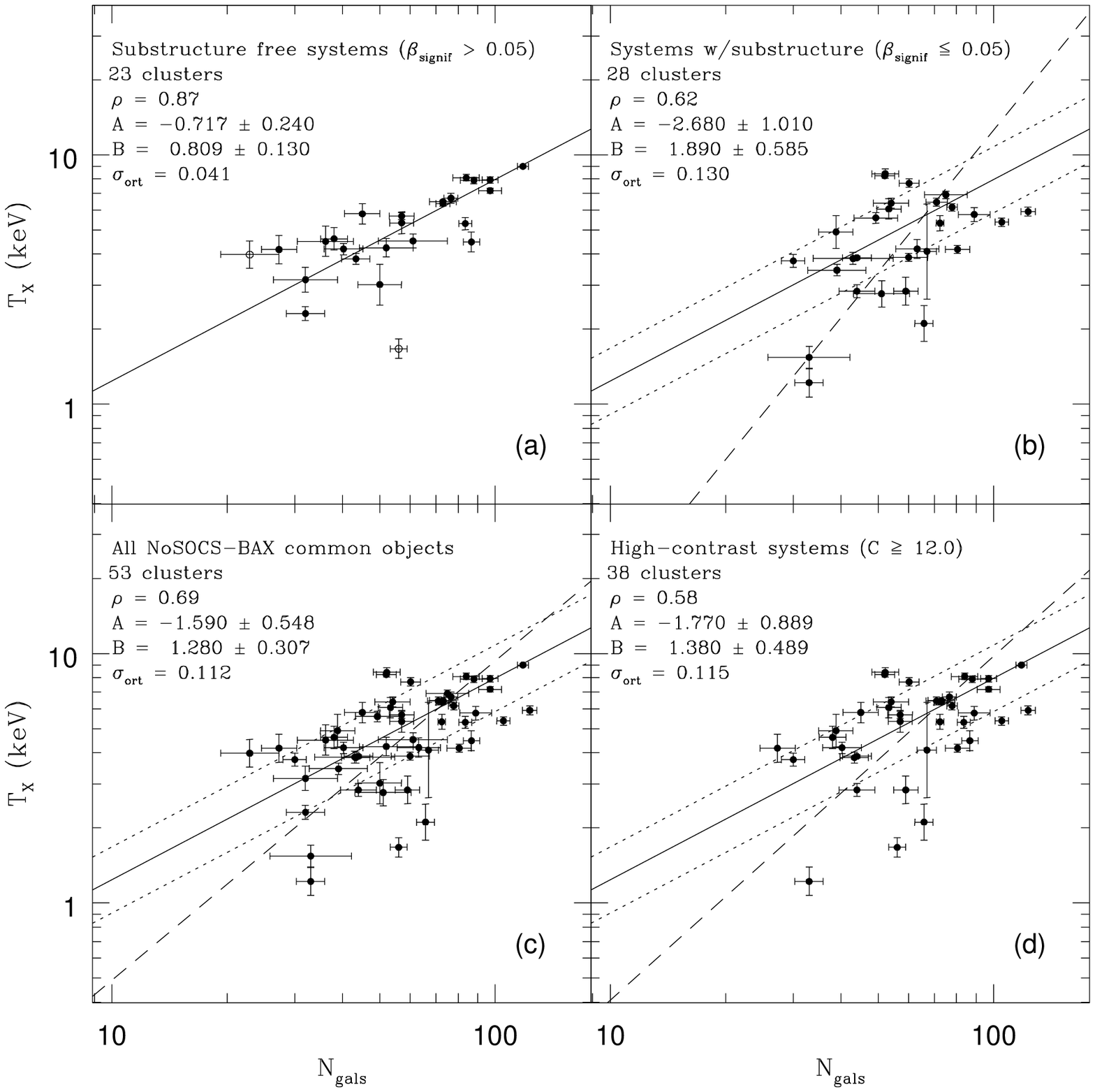}
\caption{The T$_X$-N$_{gals}$ relation for all clusters
at $0.07 \lsim z \lsim 0.21$. The full sample is shown in the lower 
left panel; high contrast systems are exhibited in the lower right panel; 
clusters considered to be substructure free are exhibited
in the upper left panel, where two outliers are
shown with an open symbol; clusters with substructure are shown in 
the upper right panel.\label{fig14}}
\end{figure*}

\subsubsection{X-ray Luminosity and Richness: Sample Dependence}

We now compare our findings to literature results. We show in Figure
17 the connection between L$_X$ and richness or L$_{opt}$. Our results
are shown by thin lines, while the literature ones are thick lines.
When performing this comparison we investigate a few effects: (i) how
the slope of the optical {\it vs} X-ray relation changes with cluster
contrast; (ii) the impact of the centroid (optical or X-ray) on the
slope; (iii) the slope difference when using richness or L$_{opt}$ as
the optical parameter; (iv) the difference in the results obtained by
the bisector and orthogonal fits. All of the relations based on our
data shown here include all clusters (we do not exclude clusters with
N$_{gals} < 10.0$, the only exception being the relation for clusters
with no substructure) and are obtained using an orthogonal
regression. We also stress that the relations based on the
X-ray center (BAX) are derived from all BAX clusters within the area
covered by NoSOCS (whether or not they have a counterpart in
DPOSS). The relations obtained with the optical center include only clusters
common to the optical and X-ray catalogs.  Hence, the former
relations are based on larger samples of clusters.

Figure 17 shows the L$_X-$N$_{gals}$ relations for a variety of samples. All of the results from our sample are plotted as thin lines, including:
\begin{itemize}
\item Solid line: all BAX clusters in the NoSOCS region. Of the 792
BAX clusters, 744 have both richness and L$_X$ available. Ten clusters
are eliminated as outliers after a 3$-\sigma$ clipping is applied.  
\item Dotted line: The previous sample trimmed to clusters with optical
contrast $\ge 5.0$.
\item Short dashed line: The sample is further reduced to clusters with
contrast $\ge 10.0$. For the above three relations we use the X-ray centroid when estimating N$_{gals}$.
\item Long dashed line: As the solid line above,
but using only the common clusters between NoSOCS and BAX (with
available L$_X$ and N$_{gals}$), and the optical centroid for
the richness calculation. 
\item Dashed$-$dotted line: The sample is then cut to consider only
substructure free systems, using the optical centroid (as shown in Figure 15). 
\item Long dashed$-$dotted line:  The L$_X-$L$_{opt}$ relation, using the X-ray centroids.
\end{itemize}
We also show a variety of results from the literature as thick lines:
\begin{itemize}
\item Solid line: The L$_X$ -- L$_{opt,i}$ relation obtained by
\citet{pop04} for their full sample. We note that they too use the X-ray centers.
\item Dotted line: The relation from \citet{pop04} when groups are excluded.
\item Short dashed line: The correlation between L$_X$ (bolometric) and optical
richness (given by $\Lambda_{cl}$) from \citet{don01}.
\item Long dashed line: The correlation between
L$_X$ and richness (B$_{gc}$) obtained by
\citet{yee03}.
\end{itemize}
We note that \citet{yee03} use a bisector solution, which is
generally less steep. If we considered a bisector solution in our
analysis the powers in the relations we obtained (thin lines) would
range from 1.79 to 2.19, which is in good agreement with their results.

From this figure we can see that the slope increases rapidly as we
consider higher contrast systems. This trend was already evident in
Figure 15, where we also see the scatter is greatly reduced when the sample
is restricted. However, this behavior does not persist for much larger
values of the contrast cut. In Figure 17 we show the results for the
full sample and with cuts at $C \ge 5.0$ and $10.0$. If we increase
the cut to $C \ge 15.0$ or $20.0$ the slope variation ceases, with the
slope actually becoming lower. That is due to sampling effects as the
number of clusters is drastically reduced. It is important to mention
that contrast is directly proportional to richness. Thus, if richness
cuts were applied -- instead of contrast cuts -- we would find similar
trends.  An important issue is that when using low luminosity or
richness systems the data is likely to be censored. \citet{don01}
extensively discuss this problem and show how the correlation between
L$_X$ and richness ($\Lambda_{cl}$) changes when the fit is derived in
conjunction with the $\Lambda_{cl}$ function and the X-ray luminosity
function ({\it i.e.}, taking into account sampling effects at low flux
regimes).

\begin{figure*}[!ht]
\centering
\includegraphics[totalheight=4.8in, scale=0.6]{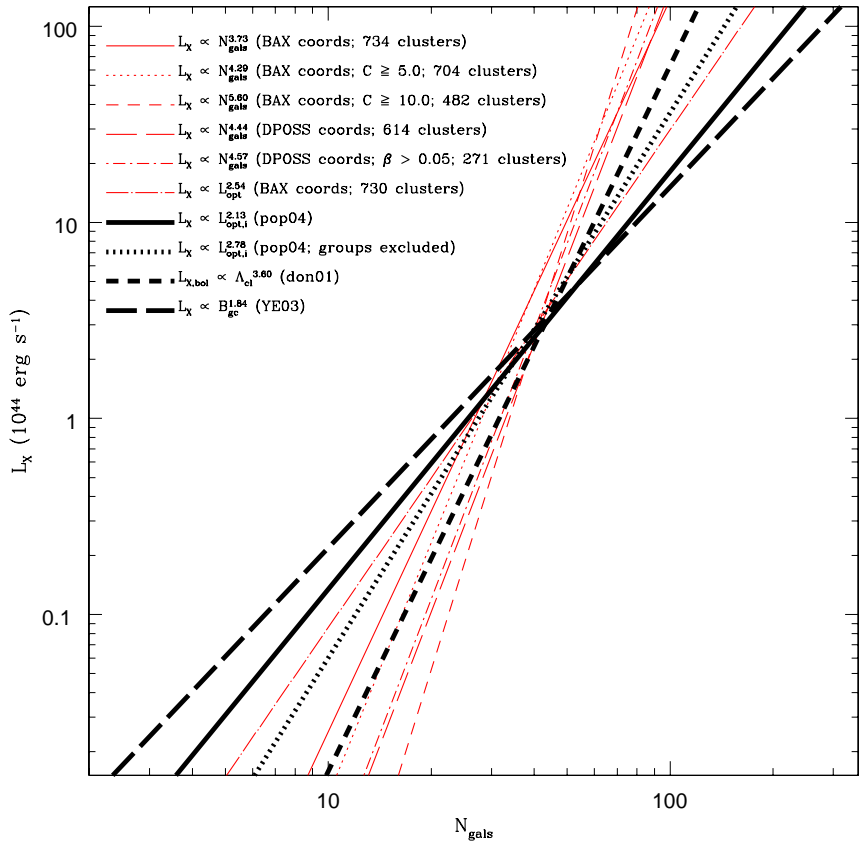}
\caption{The correlation between L$_X$ and richness or L$_{opt}$.  The
thin lines show the results obtained in this work, while the thick
lines exhibit results from the literature. This work: {\bf(i) solid line:}
L$_X-$N$_{gals}$ for all BAX clusters in the NoSOCS region; {\bf(ii)
dotted line:} the previous sample trimmed at contrast $\ge 5.0$; {\bf(iii)
short dashed line:} the sample is further cut at contrast $\ge
10.0$. For relations i--iii we considered the X-ray centroid when
estimating N$_{gals}$; {\bf(iv) long dashed line:} as (i) but considering
all common clusters between NoSOCS and BAX and using the optical
centroid for richness calculation; {\bf(v) short dashed$-$dotted:} sample
(iv) cut to consider only substructure free systems; {\bf(vi) long
dashed$-$dotted line:} The connection between L$_X$ and L$_{opt}$ (BAX
coordinates are used for computing L$_{opt}$) for all BAX systems in
the NoSOCS region. Literature results: {\bf(i) solid line:} The L$_X$ -- L$_{opt,i}$ relation
obtained by \citep{pop04} for their full sample; {\bf(ii) dotted line:} The
relation obtained by the same authors when excluding groups; {\bf(iii)
short dashed line:} L$_X$ (bolometric) {\it vs.} optical richness
($\Lambda_{cl}$) from \citet{don01}; {\bf(iv) long dashed line:} the
connection between L$_X$ and richness (B$_{gc}$) as obtained by
\citet{yee03}.
\label{fig13_02}}
\end{figure*}

Based on Figure 17, a very interesting conclusion is that the
relations are less steep when using the X-ray centroid (instead of
optical) to calculate richness (or L$_{opt}$). When using all clusters
and considering the X-ray center, the slope is 3.73, increasing to
4.29 and 5.60 when trimming the sample at $C \ge 5.0$ or $10.0$,
respectively. The slopes measured when considering the optical
centroid for these three cases (using the common BAX and NoSOCS
clusters) are 4.54, 5.09 and 6.40. This effect is associated to the
contrast cuts, as the galaxy number density is generally reduced when
we consider the X-ray centroid, in which case we expect the slopes to
be smaller. We also find that the relation using optical luminosity is
shallower than the one with richness. When comparing our results to
the literature we note that \citet{pop04} found the same effect
regarding the contrast cut (by excluding groups from their main
sample). The relations we obtain are also in good agreement to their
results, as well as those from \citet{don01} and \citet{yee03}. In the
latter case there is concordance if we consider only the bisector
solution.

We define N$_{gals}$ differently from the richness parameters in the
literature shown here ($\Lambda_{cl}$ and B$_{gc}$). Thus, the most
direct comparison of our results to the literature is made when
considering the relations involving L$_{opt}$. The results from
\citet{pop04} show less power than ours (2.13 in comparison to
2.54). However, their sample may be composed of a larger number of low
luminosity systems (groups) than ours. When they exclude groups, the
power increase to 2.78, in much better agreement with our
findings. Even considering their full sample our results agree at the
2$\sigma$ level, while the agreement is at the 1$\sigma$ level for
their high contrast sample. The use of the optical
center (instead of the X-ray centroid) for computing L$_{opt}$ yields
steeper relations.  When using all the common BAX-NoSOCS clusters we
find a slope of 3.01 for the L$_{X}$-L$_{opt}$ relation, which
disagrees with the full sample of \citet{pop04} at the 3$\sigma$
level, but is in concordance -- at the 2$\sigma$ level -- to their results 
excluding groups.
 
Another effect that merits attention is the fitting method employed
(see \S6.3).  Literature results are often contradictory, and one of
the causes that is rarely discussed is the linear regression
employed. We have shown (Figure 14) that our results differ from
\citet{mar98}, mostly because they adopt a bisector solution when
fitting the L$_{X}$-T$_{X}$ relation. On the other hand our findings
are consistent with those of \citet{dav93}. Similar differences are
shown when comparing X-ray luminosity and richness. The smaller power
obtained by \citet{yee03} in comparison to ours is mainly associated
with the fitting method (bisector in their case). It is necessary to
consider all of the possible effects (centroid adopted, flux limit and
fitting method, at a minimum) when attempting to calibrate a
mass-observable relation.  Systematic errors in the measurement of
cluster mass are exponentially amplified by the steepness of the
cluster mass function. In future work we will link richness
to mass and then quantify the effects of different solutions on the
mass function estimates.

\subsubsection{X-ray Temperature and Richness: Sample Dependence}

In Figure 18 we show the connection between T$_X$ and richness or
L$_{opt}$. As before, our findings are shown by thin lines, while
thick lines show the literature results. We have also not excluded
clusters with N$_{gals} < 10.0$ (except for the relation of
clusters without substructure), and the fits are obtained through
orthogonal regression. The results shown here include only clusters at
$0.07 \lsim z \lsim 0.21$. The thin lines (our work) are analogous to
those in the previous figure, but for the relations involving
T$_X$. The literature results (thick lines) are the T$_X-$L$_{opt,i}$
relation obtained by \citet{pop04} (solid line), the T$_X-$B$_{gc}$
relation from \citet{yee03} (dotted line), the T$_X-$L$_{NIR}$
relation from \citet{oh06} (short dashed line) and the 
T$_X-$L$_{opt,r}$ relation obtained by \citet{pli05} (long dashed line).
We do not find the same
trend with contrast as for L$_X-$N$_{gals}$. This may be due to the
smaller fraction of clusters excluded based on contrast for
the T$_X-$N$_{gals}$ relation (see above), such that this relation is
already restricted to high contrast systems. There is still a small
difference between the results with different centroids, and L$_{opt}$
again defines a shallower relation than N$_{gals}$. Our findings also
show good agreement with the literature results.

\begin{figure*}[!ht]
\centering
\includegraphics[totalheight=4.8in, scale=0.6]{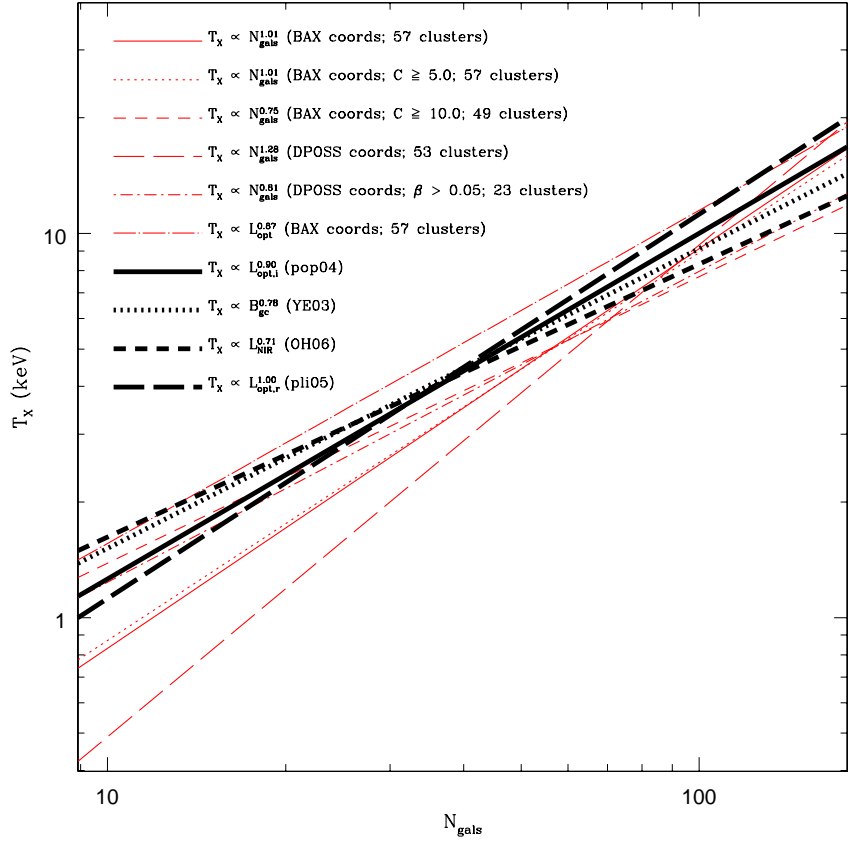}
\caption{The connection between temperature and richness or optical
luminosity. The thin lines show the results obtained in this work,
while the thick lines exhibit results from the literature. The meaning
of the thin lines is analogous to the previous figure, but now for the
T$_X-$N$_{gals}$ relation and only considering clusters at $0.07 \lsim
z \lsim 0.21$. The literature results (thick lines) are given for: {\bf(i) solid
line:} the connection between T$_X$ and L$_{opt,i}$ obtained by
\citet{pop04}; {\bf(ii) dotted line:} the T$_X-$B$_{gc}$ relation
obtained by \citet{yee03}; {\bf(iii) short dashed line:} the
T$_X-$L$_{NIR}$ correlation from \citet{oh06}; and 
{\bf(iv) long dashed line:} the T$_X-$L$_{opt,r}$ correlation 
from \citet{pli05}.
\label{fig13_03}}
\end{figure*}

A few comments on our results are still necessary. If it is assumed
that mass traces optical light and the gas is in hydrostatic
equilibrium, we expect the mass-luminosity ($M/L_{opt}$) ratio to be
constant and the gas temperature to be proportional to mass ($T
\propto M^{2/3}$). The X-ray luminosity is also related to temperature
($L_X \propto T^{\alpha}$), although the precise value of ${\alpha}$
is subject to some debate. The theoretical expectation is ${\alpha} =
2$, but its observed estimate is closer to 3 (this work and
\citet{dav93}). Considering these relations and assuming ${\alpha} =
3$, for instance, we expect the X-ray and optical luminosities to be
related as ($L_X \propto L_{opt}^2$) and the relation to temperature
to be $T_X \propto L_{opt}^{0.7}$. If we consider that N$_{gals}$ is
directly proportional to L$_{opt}$ our results point to steeper
relations.  $L_X \propto L_{opt}^3$ and $T_X \propto L_{opt}^{0.8}$,
although $T_X-$L$_{opt}$ becomes less steep if we exclude clusters
with substructure. As pointed out by \citet{don01} and \citet{pop04},
the steepness of the X-ray to optical relations point to the fact that
the mass-luminosity ($M/L_{opt}$) ratio of galaxy clusters is not
likely to be constant. This is actually demonstrated by \citet{pop05},
who found $M/L \propto M^{0.2}$. We note that this conclusion could
not be made if we had considered the bisector solution. Had we used
the bisector, we would find $L_X \propto L_{opt}^{1.80}$, which is
consistent with a constant mass-luminosity ratio. In a future paper we
plan to investigate in detail this relation ($M/L_{opt}$) and its
possible dependence on substructure in clusters.

\section{Summary}

We have compared a list of X-ray galaxy clusters selected from the
literature (BAX) to an optical galaxy cluster catalog (NoSOCS).  This
comparison covers the whole Northern Hemisphere for $|b| > 30^\circ$
and spans the redshift range $0.05 \le z \le 0.40$. The X-ray and
optical centroids show excellent agreement, with a typical
X-ray$-$optical offset of $< 0.50$ $h^{-1}$ Mpc (or $< 400''$). The
overall recovery rate of X-ray clusters in the optical is $81\%$, with
the missing clusters typically poor and/or distant. Thus, NoSOCS 
efficiently recovers nearly all X-ray luminous clusters from BAX
(L$_X \gsim 3.2$ $10^{44}$ $ergs$ $s^{-1}$) at all redshifts below $z
\sim 0.2$. In terms of richness, the recovery rate is $90\%$ for
clusters with N$_{gals} \gsim 25$ out to $z \sim 0.2$, and for all clusters
with N$_{gals} \gsim 80$ out to $z \sim 0.3$.

We employed four statistical tests to search for substructure using
optical imaging data. These tests were optimized through the
minimization of the scatter in the relations between optical and X-ray
properties (namely T$_X-$N$_{gals}$). We have also shown the
dependence of richness on the centroid and aperture. We investigated
the dependence of substructure results on these parameters, as well as
on the cluster contrast, richness, redshift and magnitude range used
for sampling galaxies. The
substructure results shown here are based on the largest sample used
for this purpose to date (10190 clusters). As noted by \citet{pin96}
some tests are much more sensitive than others. For instance, the
$\beta$, or symmetry, test indicates that 35$\%$ of the clusters have
substructure at the 5$\%$ significance level (in a radius of $
1.50$ $h^{-1}$ Mpc and magnitude range of $m^*_r-1 \le m_r \le m^*_r+1$),
while the FE test finds $21\%$, the Lee 2D test indicates $45\%$ and
AST only $13\%$.

We have also compared richness (N$_{gals}$) and X-ray cluster properties
(L$_X$ and T$_X$). We find that T$_X$ correlates better to
N$_{gals}$ than L$_X$. We examined the potentially dominant factors 
affecting these scaling relations. We find that the correlations are most 
sensitive to the aperture used for estimating richness. For 
larger radii we sample many more cluster galaxies, but the background 
noise also becomes much stronger. We find that the optimal radius is 
$ 0.50$ $h^{-1}$ Mpc, which is in good agreement to that found by 
\citet{pop04}. Adopting this radius for estimating N$_{gals}$, we 
find that the cluster contrast is the dominant source of scatter 
in the L$_X-$N$_{gals}$ relation. The presence of substructure does 
not affect the scatter of this relation. On the other 
hand, when comparing T$_X$ to the optical parameters we find that 
substructure has a strong effect on the scatter.

Our findings also corroborate previous results in the literature,
although we note some issues that could explain some of the
difficulties in performing these comparisons. These are mainly related
to the contrast of the systems, as higher contrast systems generally
define steeper relations. For instance, if we compare samples that
span different mass regimes we may find very different correlations,
as the contrast cuts will be very different. Relations derived from
richness (or L$_{opt}$) calculated from the X-ray centroid have
smaller slopes than ones obtained using the optical centroid, due to
the smaller galaxy number density seen around the X-ray centroids. The
use of different linear regression methods can also lead to
inconsistent results, especially for very steep and noisy relations.
As shown in Section 6, the bisector solution is generally less
efficient in recovering the true slopes. Literature results would likely
show much better agreement if the same linear regression
methods were used. The steeper slopes obtained here also
suggest (as have other recent works) that the
mass-luminosity ratio may not be constant. The 
fitting method and centroid used are crucial for reaching
this conclusion. For the L$_X-$L$_{opt}$ relation the bisector
solution indicates $L_X \propto L_{opt}^{1.80}$, which is close to the
expectations from a constant mass-luminosity ratio.

In the future we plan to use both deeper optical and X-ray data to
investigate substructure. These will allow us to assess the
reliability of the optical estimates, as well as to investigate the
evolution of substructure with look-back time. We also plan to use
apertures that scales with mass, such as $R_{500}$ and $R_{200}$ to
see if the structure segregation we show here is also found with these
apertures. Such virial radii are more typically used when measuring
cluster parameters from large-scale simulations, and it is crucial to
use the same techniques in comparing these with observations. Finally,
we plan to investigate the mass-luminosity ratio, as well as estimate
the cluster mass function with the NoSOCS data.

\acknowledgments

The processing of DPOSS and the production of the Palomar-Norris Sky
Catalog (PNSC) on which this work was based was supported by generous
grants from the Norris Foundation, and other private donors. Some of
the software development was supported by the NASA AISRP program.  We
also thank the staff of Palomar Observatory for their expert
assistance in the course of many observing runs. We would like to
thank the anonymous referee for useful comments. Finally, we
acknowledge the efforts of the POSS-II team, and the plate scanning
team at STScI. PAAL was supported by the Funda\c c\~ao de Amparo
\`a Pesquisa do Estado de S\~ao Paulo (FAPESP, process 03/04110-3).
SGD and AAM were supported in part by the NSF grant AST-0407448.
Several undergraduates participated in the data acquisition and
processing toward the photometric calibration of DPOSS. PAAL would
like to thank Jason Pinkney for sharing the code used for estimating
substructure and for helpful discussions on this code's implementation.
This research has made use of the X-Rays Clusters Database (BAX)
which is operated by the Laboratoire d'Astrophysique de Tarbes-Toulouse 
(LATT), under contract with the Centre National d'Etudes Spatiales (CNES).



\begin{thebibliography}{}

\bibitem[Akritas and Bershady(1996)]{akr96}  Akritas, M., Bershady, M. 1996,
\apj, 470, 706

\bibitem[Bahcall et al.(1997)]{bah97} Bahcall, N.\ A., Fan, X., 
Cen, R.\ 1997, \apj, 485, L53

\bibitem[Bahcall et al.(2003)]{bah03} Bahcall, N.\ A., Dong, F., 
Bode, P., Kim, R. \etal 2003, \apj, 585, 182

\bibitem[Basilakos et al.(2004)]{bas04} Basilakos, S., Plionis, M., 
Georgakakis, A., Georgantopoulos, I. \etal 2004, \mnras, 351, 989

\bibitem[Blanchard and Bartlett(1998)]{bla98} Blanchard, A. and 
Bartlett, J. G. 1998, \aap, 332, 49

\bibitem[Carlberg et al.(1997)]{car97} Carlberg, R., Morris, S.,
Yee, H., Ellingson, E.\ 1997, \apj, 479, L19

\bibitem[Coleman et al.(1980)]{col80} Coleman, G. D., Wu, C. C., Weedman, 
D. W. 1980, \apjs, 43, 393

\bibitem[David et al.(1993)]{dav93} David, L. P., Slyz, A., Jones, C.
Forman, W., Vrtilek, S. D., Arnaud, K. A. 1993, \apj, 412, 479

\bibitem[Donahue et al.(1998)]{don98} Donahue, M., Voit, G., Gioia, I., 
Lupino, G., Hughes, J., Stocke, J. 1998, \apj, 502, 550

\bibitem[Donahue et al.(2001)]{don01}Donahue, M., Mack, J., Scharf, C.
 \etal 2001, \apj, 552, L93

\bibitem[Donahue et al.(2002)]{don02}Donahue, M., Scharf, C., Mack, J.,
\etal 2002, \apj, 569, 689

\bibitem[Eke et al.(1998)]{eke98} Eke, V., Cole, S., Frenk, C. and 
Henry,J. 1998, \mnras, 298, 1145

\bibitem[Feigelson et al.(1992)]{fei92} Feigelson, E. D. And Babu, G. J. 1992,
\apj, 397, 55

\bibitem[Fitchett \& Webster(1987)]{fit87} Fitchett, M. J. and Webster, R.
1987, \apj, 317, 653

\bibitem[Fitchett(1988)]{fit88} Fitchett, 1988, \mnras, 230, 161

\bibitem[Gal et al.(2003)]{gal03} Gal, R.\ R., de Carvalho, R. R., 
Lopes, P.\ A.\ A., Djorgovski, S.\ G., Brunner, R.\ J., Mahabal, A.\ A., 
Odewahn, S.\ C.\ 2003, \aj, 125, 2064

\bibitem[Gal et al.(2006)]{gal06} Gal, R.\ R., de Carvalho, R. R., 
Lopes, P.\ A.\ A., Djorgovski, S.\ G., Brunner, R.\ J., Mahabal, A.\ A., 
Odewahn, S.\ C.\ 2005, \aj, {\it in prep.}

\bibitem[Gilbank(2001)]{gil01} Gilbank, D.\ 2001, Ph.D.\ Thesis, University
of Durham

\bibitem[Gilbank et al.(2004)]{gil04} Gilbank, D. G., Bower, R., 
Castander, F., Ziegler, B. L. 2004, \mnras, 348, 551

\bibitem[Isobe et al.(1990)]{iso90} Isobe, T., Feigelson, E. D., 
Akritas, M. and Babu, G. J. 1990, \apj, 364, 104

\bibitem[Jones \& Forman(1999)]{jon99} Jones, C., \& Forman, W. 1999,
\apj, 511, 65

\bibitem[Kitayama and Suto(1997)]{kit97}Kitayama, T. and Suto, Y. 
1997, \apj, 490, 557

\bibitem[Kolokotronis et al.(2001)]{kol01} Kolokotronis, V., Basilakos, S.,
Plionis, M. and Georgantopoulos, I. 2001, \mnras, 320, 49

\bibitem[Kriessler \& Beers(1997)]{kri97} Kriessler, J. R. and Beers, T. C.
1997, \aj, 113, 80

\bibitem[Lee(1979)]{lee79} Lee, K. L. 1979, J. Am. Stat. Assoc., 74 
(367), 708

\bibitem[L\'opez-Cruz et al.(2004)]{lop04} L\'opez-Cruz, O., Barkhouse, W.,
Yee, H., \apj, 614, 679

\bibitem[Lubin et al.(2004)]{lub04} Lubin, L.\ M., Mulchaey, J.\ S. \& Postman, M.\ 2004, \apj, 601, L9  

\bibitem[Markevitch(1998)]{mar98} Markevitch, M. 1998, \apj, 504, 27

\bibitem[Mathiesen and Evrard(1998)]{mat98} Mathiesen, B. 
and Evrard, A. E. 1998, \mnras, 295, 76

\bibitem[Mathiesen, Evrard and Mohr(1999)]{mat99} Mathiesen, B., 
Evrard, A. E., Mohr, J. J. 1998, \apj, 520, L21

\bibitem[Mohr et al.(1995)]{moh95} Mohr, J. J., Evrard, A. E., 
Fabricant, D. G., Geller, M. J. 1995, \apj, 447, 8

\bibitem[O'Hara et al.(2006)]{oh06} O'Hara, T. B., Mohr, J. J., Bialek, J. J.,
Evrard, A. E. 2006, \apj, 639, 64

\bibitem[Paolillo et al.(2001)]{pao01} Paolillo, M., Andreon,
S., Longo, G., Puddu, E., Gal, R.~R., Scaramella, R., Djorgovski, S.~G., \&
de Carvalho, R.\ 2001, \aap, 367, 59

\bibitem[Pinkney et al.(1996)]{pin96} Pinkney, J., Roettiger, K., Burns, J.,
Bird, C. 1996, \apjs, 104, 1

\bibitem[Plionis et al.(2005)]{pli05} Plionis, M., Basilakos, S., 
Georgantopoulos, I., Georgakakis, A. 2005, \apj, 622, L17

\bibitem[Popesso et al.(2004)]{pop04} Popesso, P., B\"ohringer, H., 
Brinkmann, J., Voges, W., York, D. 2004, \aap, 423, 449

\bibitem[Popesso et al.(2005)]{pop05} Popesso, P., Biviano, A., 
B\"ohringer, H., Romaniello, M., Voges, W. 2005, \aap, 433, 431

\bibitem[Postman, Lauer, Oegerle, \& Donahue(2002)]{pos02} Postman, M.,
Lauer, T.~R., Oegerle, W., \& Donahue, M.\ 2002, \apj, 579, 93

\bibitem[Reichart et al.(1999)]{rei99} Reichart, D. E., Nichol, R. C., 
Castander, F. J. \etal 1999, \apj, 518, 521

\bibitem[Rhee, van Haarlem and Katgert(1991)]{rhe91} Rhee, G. F. R. N., 
van Haarlem, M. P. and Katgert, P. 1991, \aaps, 91, 513

\bibitem[Rhee, van Haarlem and Katgert(1991b)]{rhe91b} Rhee, G. F. R. N., 
van Haarlem, M. P. and Katgert, P. 1991b, \aap, 246, 301

\bibitem[Rines \& Diaferio(2006)]{rin06} Rines, K., Diaferio, A.  2006, 
submitted to \aj (astro-ph/0602302)

\bibitem[Smith et al.(2005)]{smi05} Smith, G. P., Kneib, Jean-Paul, 
Smail, I., Mazzotta, P., Ebeling, H., Czoske, O. 2005, \mnras, 359, 417

\bibitem[Voit(2005)]{voi05} Voit, G. M. 2005, Rev. Mod. Phys., 77, 207

\bibitem[West, Oemler and Dekel(1988)]{wes88} West, M. J., Oemler, A. 
and Dekel, A. 1988, \apj, 327, 1

\bibitem[Yee and Ellingson(2003)]{yee03} Yee, H. K. C. and 
Ellingson, E. 2003, \apj, 585, 215

\end{thebibliography}
\end{document}